\documentclass[twocolumn,aps,showpacs,nofootinbib,prd]{revtex4-1}
\usepackage{amssymb}
\usepackage{amsmath}
\usepackage{graphics}
\usepackage{epsfig}
\usepackage[dvips]{color}

\newcommand{\T}{\tilde}
\newcommand{\n}{\mathbf}
\newcommand{\be}{\begin{equation}}
\newcommand{\ee}{\end{equation}}
\newcommand{\bea}{\begin{eqnarray}}
\newcommand{\eea}{\end{eqnarray}}
\newcommand{\beas}{\begin{eqnarray*}}
\newcommand{\eeas}{\end{eqnarray*}}

\newcommand{\nn}{\nonumber\\}

\newcommand{\bd}[1]{{\bf #1}}

\begin{document}
\title{Impact of the energy loss spatial profile and shear viscosity to entropy density ratio for the Mach cone vs. head shock signals produced by a fast moving parton in a quark-gluon plasma}
\author{Alejandro
Ayala$^{1,4}$, Jorge David Casta\~no-Yepes$^1$, Isabel Dominguez$^2$ and Maria Elena
  Tejeda-Yeomans$^{3,1}$}
  \address{
  $^1$Instituto de Ciencias
  Nucleares, Universidad Nacional Aut\'onoma de M\'exico, Apartado
  Postal 70-543, M\'exico Distrito Federal 04510,
  Mexico.\\
  $^2$Facultad de Ciencias F\'isico-Matem\'aticas, Universidad Aut\'onoma de Sinaloa,
  Avenida de las Am\'ericas y Boulevard Universitarios, Ciudad Universitaria,
  C.P. 80000, Culiac\'an, Sinaloa, M\'exico.\\
  $^3$Departamento de F\'{\i}sica,
  Universidad de Sonora, Boulevard Luis Encinas J. y Rosales, Colonia
  Centro, Hermosillo, Sonora 83000, Mexico.\\
  $^4$Centre for Theoretical and Mathematical Physics, and Department of Physics,
  University of Cape Town, Rondebosch 7700, South Africa}

\begin{abstract}

We compute the energy and momentum deposited by a fast moving parton in a quark-gluon plasma using linear viscous hydrodynamics with an energy loss per unit length profile proportional to the path length and with different values of the shear viscosity to entropy density ratio. We show that when varying these parameters, the transverse modes dominate over the longitudinal ones and thus energy and momentum is preferentially deposited along the head-shock, as in the case of a constant energy loss per unit length profile and the lowest value for the shear viscosity to entropy density ratio. 

\end{abstract}

\maketitle

\section{Introduction}\label{I}

Experiments where heavy nuclei are collided at high energies at the BNL Relativistic Heavy-Ion Collider~\cite{RHIC} and the CERN Large Hadron Collider~\cite{LHC} show that in these reactions the so-called Quark Gluon Plasma (QGP) is formed. The dynamics of the bulk matter can be accurately described using viscous hydrodynamics~\cite{viscoshydro}. One way to study the properties of the QGP is to consider how hard scattered partons transfer energy and momentum to this medium.

In a given event the particle with the largest momentum defines the near-side, and the opposite side is called the away-side. In practice, one considers that the hard scattering happens near the fireball's surface in such a way that the away-side patrons deposit energy and momentum to the medium, whereas the near-side ones fragment in vacuum, giving rise to the so-called jet-quenching~\cite{jetq1,jetq2}. A possible way to characterize the medium is to study azimuthal particle correlations.

These correlations show some interesting features: When the leading and the away-side particles have similar momenta, the correlation shows a suppression of the away-side peak, compared to proton collisions at the same energies. However, when the momentum difference between leading and away-side particles increases, either a double peak or a broadening of the away-side peak appears. Neither of these features are present in proton collisions at the same energies~\cite{azcor}.

Explanations based on the emission of sound modes caused by one fast moving parton~\cite{Casalderrey, Renk, Betz}, the so-called {\it Mach cones} are nowadays considered incomplete, since the jet-medium interaction produces also a wake whose contribution cannot be ignored~\cite{Casalderrey2, Torrieri}. Moreover it was recently shown that it is unlikely that the propagation of a single high-energy particle through the medium leads to a double-peak structure in the azimuthal correlation in a system of the size and finite viscosity relevant for heavy-ion collisions, since the energy momentum deposition in the head shock region is strongly forward peaked~\cite{Bouras}. In addition, the overlapping perturbations in very different spatial
directions wipe out any distinct Mach cone structure, according to the findings of Ref.~\cite{Neufeld3, Neufeld78}.

Currently, the origin of the  double peak/broadening is described in terms of initial state fluctuations of the matter density in the colliding nuclei. Nevertheless there is also evidence of a strong connection between the observed away-side structures and the medium's path length, expressed through the dependence of the azimuthal correlation on the trigger particle direction with respect to the event plane as measured in away-side correlation studies performed by the STAR Collaboration~\cite{pathlength}. This connection is made by observing that for selected trigger and associated particle momenta, the double peak is present (absent) for out-of-plane (in-plane) trigger particle direction. A final-state effect rather than an initial state one, seems more consistent with this observation~\cite{nostrum}.

The energy momentum transferred by the fast traveling parton to the medium can be described in terms of linearized viscous hydrodynamics~\cite{Casalderrey2, linhydro, Neufeld79, review}. An important ingredient for this description is the energy loss per unit length $dE/dx$ which enters as the coefficient describing a local hydrodynamic source term. It is known that  this parameter exhibits a non-trivial dependence on the traveled path length $L$. For instance, depending on the interplay between the evolving density of the medium during the collision, the medium's size and formation length and the dominating energy loss mechanism (radiative or collisional), $dE/dx$ could be either constant or proportional to $L$~\cite{Eloss1,lineal,Neufeld82,Neufeld103}.

In a previous work~\cite{ADY}, we have explored the consequences drawn from assuming that $dE/dx$ is constant and that the shear viscosity to entropy density ratio takes on its lower theoretical value, showing that under such scenario the Mach cone signal is weaker as compared to the wake or head-shock. Moreover, we also showed that under such conditions, the double peak/broadening in azimuthal angular correlations can be better described by two instead of one parton depositing energy and momentum into the medium. In this work we set out to explore the consequences of a linear dependence on $L$ of the energy loss per unit length and different values of the shear viscosity to entropy ratio, using the same framework. The paper is organized as follows: In Sec.~\ref{II}, we obtain the expression for the local hydrodynamic source in Fourier space and, from the solution to the linear viscous hydrodynamic equations, we obtain the energy and momentum deposited by the source into the medium. In Sec.~\ref{III} we study different allowed values for the model parameters, in particular different traveled paths and different shear viscosity to entropy density ratios. We show that the energy and momentum is still preferentially deposited along the head-shock, as in the case of a constant energy loss per unit length profile and the lowest value for the shear viscosity to entropy density ratio. We finally summarize and conclude in Sec.~{IV}.

\section{Hydrodynamical description of energy loss}\label{II}

To describe the interaction between a fast moving parton and the medium, one can resort to linearized viscous hydrodynamics. In such a description, the source of energy-momentum is provided by the current produced by the fast moving parton given by
\begin{equation}
J^{\nu}(\mathbf{x},t)=\left(\frac{dE}{dx}\right)v^{\nu}\delta^{3}(\mathbf{x}-\mathbf{v}t),
\label{2}
\end{equation}
where $v^\nu$ is the particle's four-velocity and $dE/dx$ is the energy loss per unit length. The current is proportional to the instantaneous location of the particle which is modeled by the three dimensional delta function.

We assume that the disturbance induced by the fast moving parton is small such that the energy-momentum tensor can be written as
\bea
   \Theta^{\mu\nu} = \Theta_0^{\mu\nu} + \delta \Theta^{\mu\nu}
   \label{tensor}
\eea
where $\delta \Theta^{\mu\nu}$ is the disturbance generated by the
parton and $\Theta_0^{\mu\nu}$ is the equilibrium energy-momentum
tensor of the underlying medium. The tensor's components
satisfy
\bea
   \partial_\mu \delta \Theta^{\mu\nu} &=& J^\nu\nn
   \partial_\mu \Theta_0^{\mu \nu} &=& 0,
\label{lin_source}
\eea
where $J^\nu$ is given in Eq.~(\ref{2}). Equations~(\ref{lin_source}) are solved by considering that $\Theta^{\mu\nu}$ consists of a term that describes an isotropic fluid
\bea
\Theta_0^{\mu \nu} = -p g^{\mu\nu} + (\epsilon + p)u_0^\mu u_0^\nu,
\eea
and the disturbance $\delta\Theta^{\mu\nu}$ that, to first order in the shear viscosity density $\eta$~\cite{Neufeld79} and ignoring bulk viscosity, has explicit components given by
\bea
\label{thetapert}
\delta\Theta^{00} &=& \delta \epsilon, \nonumber\\
\delta\Theta^{0i} &=& {\mathbf g}, \nonumber \\
\delta\Theta^{ij} &=& \delta_{ij}c_s^2 \delta\epsilon - \frac{3}{4}\Gamma_s(\partial^i\mathbf{g}^j + \partial^j\mathbf{g}^i -\frac{2}{3}\delta_{ij}\nabla \cdot {\mathbf g} ).
\eea
Here we have defined $\epsilon (t, \mathbf{x}) = \epsilon_0 + \delta \epsilon (t,\mathbf{x})$,
with $\epsilon_0$ the energy density of the background fluid and,
$\delta\epsilon$ and ${\mathbf{g}}$ the energy and momentum densities associated to the disturbance, respectively. The vector ${\mathbf{g}}$ is related to the spatial part of the medium's four-velocity,
\bea
{\mathbf{u}}=\frac{\mathbf{g}}{\epsilon_0(1+c_s^2)},
\label{g}
\eea
where $c_s$ is the sound velocity and
\bea
\Gamma_s\equiv \frac{4}{3}\frac{\eta}{\epsilon_0(1+c_s^2)}=\frac{4}{3}\frac{\eta}{s_0T}
\label{defGammas}
\eea
is the sound attenuation length, with $s_0$ the entropy density and $T_0$ the temperature of the underlying medium.

For the linear approximation the dynamical description of the disturbance is given by the first of Eqs.~(\ref{lin_source}), whose explicit components can be written
as
\bea
\label{beforefourier}
\partial_0 \delta\epsilon + \nabla\cdot\mathbf{g} &=& J^0, \nonumber \\
\partial_0 \bd{g}^i + \partial_j \delta\Theta^{ij} &=& J^i.
\eea
These equations can be readily solved in momentum space. We define
the Fourier transform pair $f({\mathbf{x}},t)$ and
$f({\mathbf{k}},\omega)$ as
\bea
f({\mathbf{x}},t) = \frac{1}{(2 \pi)^4}\int d^3 k \int d \omega
\,e^{i \bd{k}\cdot\bd{x} - i \omega t} f({\mathbf{k}},\omega).
\label{FT}
\eea

Using Eq.~(\ref{FT}) into Eqs.~(\ref{beforefourier}), together with Eqs.~(\ref{thetapert}),
we obtain
\bea
\label{afterfourier}
-i\omega \delta\epsilon + i \mathbf{k}\cdot\mathbf{g} &=& J^0, \nonumber \\
-i\omega\mathbf{g}^i + ic_s^2k^i\delta\epsilon + \frac{3}{4}\Gamma_s(k^2\mathbf{g}^i
+ \frac{k^i}{3}(\mathbf{k}\cdot\mathbf{g})) &=& J^i.
\eea

If we decompose ${\mathbf{g}}$ into its longitudinal and transverse parts, with respect to the Fourier mode ${\mathbf{k}}$, in the form
\bea
   {\mathbf{g}} = {\mathbf{g}}_L + {\mathbf{g}}_T,
\label{translong}
\eea
with the definition of longitudinal and transverse components of any
vector $\mbox{\boldmath${\sigma}$}$ given by
\bea
   \mbox{\boldmath${\sigma}$}_L&\equiv&\frac{(\mbox{\boldmath${\sigma}$}
   \cdot{\mathbf{k}})}{k^2}{\mathbf{k}},\\
   \mbox{\boldmath${\sigma}$}_T&\equiv&\mbox{\boldmath${\sigma}$} - \mbox{\boldmath${\sigma}$}_L,
\label{expltranslong}
\eea
we can solve Eqs.~(\ref{afterfourier}) for each of the ${\mathbf{g}}$ modes as well as for the energy density $\delta\epsilon$, which gives
\bea
   \delta\epsilon
   ({\mathbf{k}},\omega) &=& \frac{i {\mathbf{k}}\cdot{\mathbf{J}}({\mathbf{k}},\omega) +
   J^0({\mathbf{k}},\omega)(i \omega - \Gamma_s k^2)}{\omega^2 - c_s^2
   k^2 + i \Gamma_s \omega k^2},
\label{eps}\\
   {\mathbf{g}}_L ({\mathbf{k}},\omega) &=& \frac{i\left[\frac{\omega}{k^2}
   {\mathbf{k}}\cdot{\mathbf{J}}({\mathbf{k}},\omega)+ c_s^2 J^0({\mathbf{k}},\omega)\right]{\mathbf{k}}}
   {\omega^2 - c_s^2 k^2 + i \Gamma_s \omega k^2},
\label{gl}\\
    \bd{g}_T({\mathbf k},\omega) &=& \bd{g} - \bd{g}_L = \frac{i{\mathbf
    J}_T({\mathbf k},\omega)}{\omega + i \frac{3}{4}\Gamma_s k^2}.
\label{gt}
\eea
In a recent study~\cite{ADY}, $dE/dx$ was taken as constant. Under such assumption it was found that the longitudinal signal is weaker than the transverse one and that since the former is mostly directed along the perpendicular direction of motion of the source whereas the latter is forward peaked, the energy-momentum was preferentially deposited along the direction of motion of the hard parton. Nevertheless, it is known that depending on the size and treatment of the scattering properties of the medium, $dE/dx$ can depend on the traveled path. Let us therefore consider a simple scenario where the length dependence of $dE/dx$ is linear, namely let us take
\bea
    \left(\frac{dE}{dx}\right)=Cz,
\label{newdEdx}
\eea
where we have explicitly considered that the particle's direction of motion is along the $\hat{z}$ direction and introduced the dimensionful proportionality constant $C$ which is fixed later on. We thus write explicitly the current as
\bea
   J^{\nu}(\mathbf{x},t)=Czv^{\nu}\delta^{3}(\mathbf{x}-\mathbf{v}t),
   \label{Jexpl}
\eea
whose Fourier transform can be written as
\begin{eqnarray}
   J^{\nu}(\mathbf{k},\omega)=-2iCv\pi v^\nu\frac{\partial}
   {\partial\omega}\delta(\omega-\mathbf{k}\cdot\mathbf{v})
   \label{3}.
\end{eqnarray}
\begin{figure*}[t]
{\centering
{\includegraphics[scale=0.4]{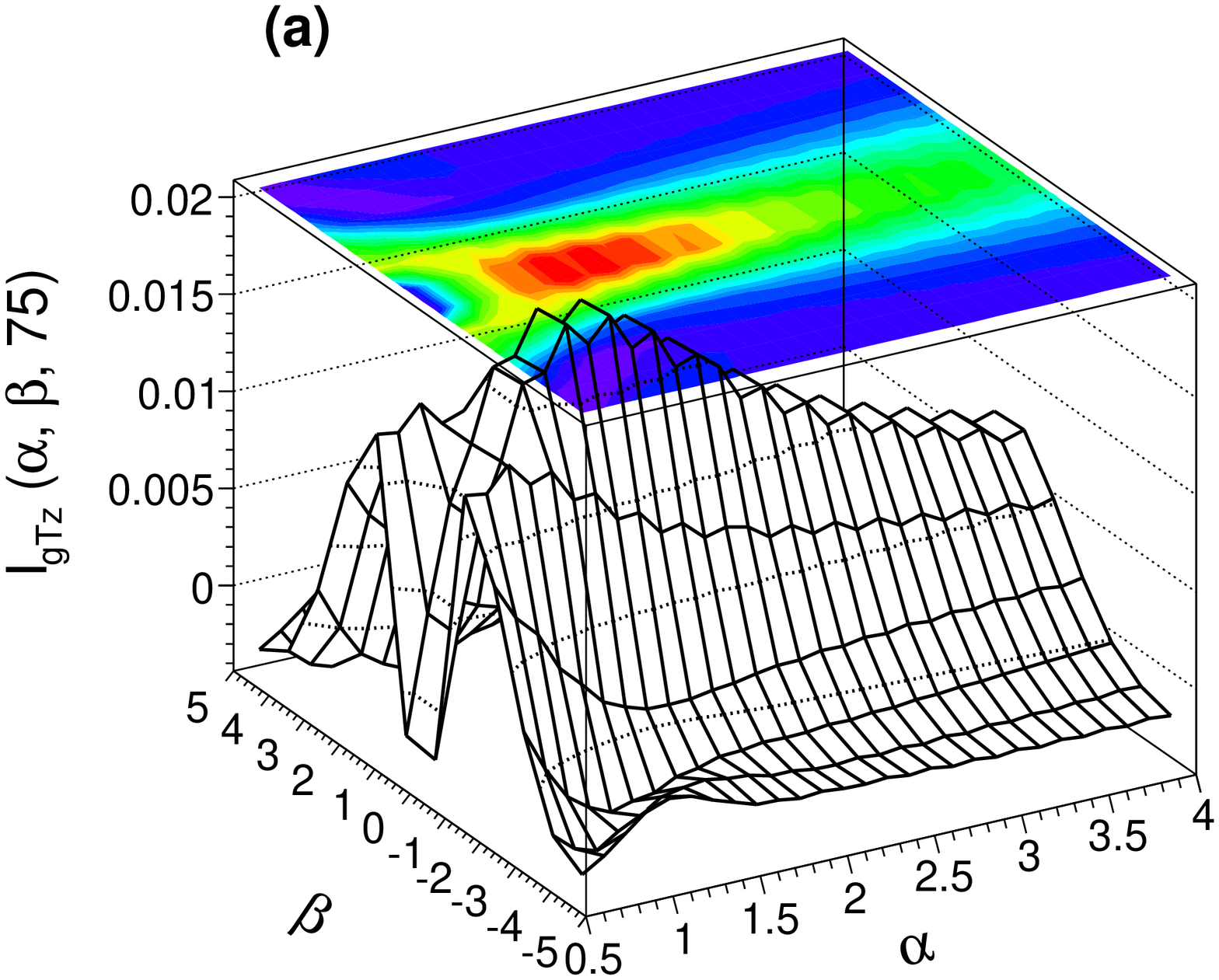}\includegraphics[scale=0.4]{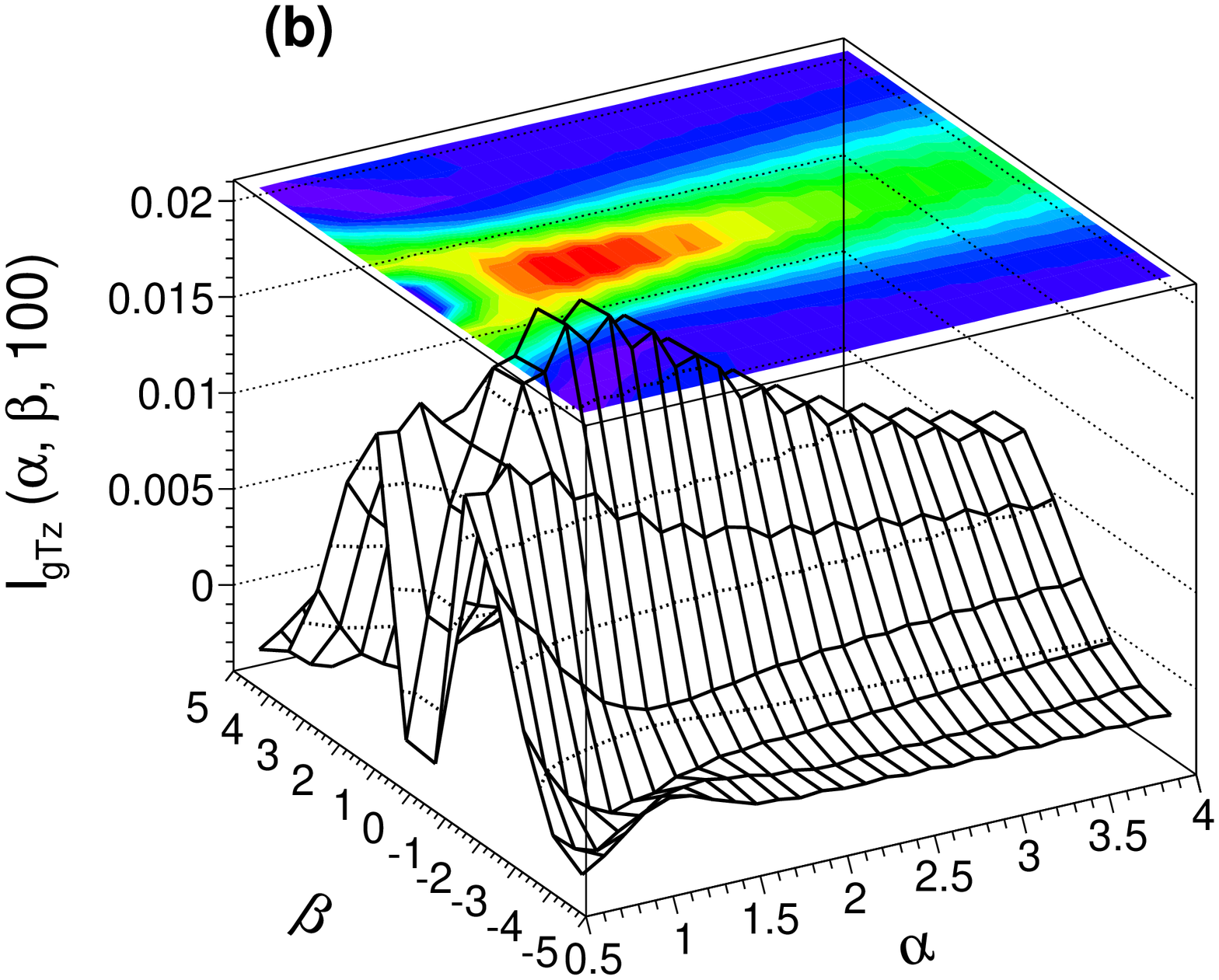}
\\\includegraphics[scale=0.4]{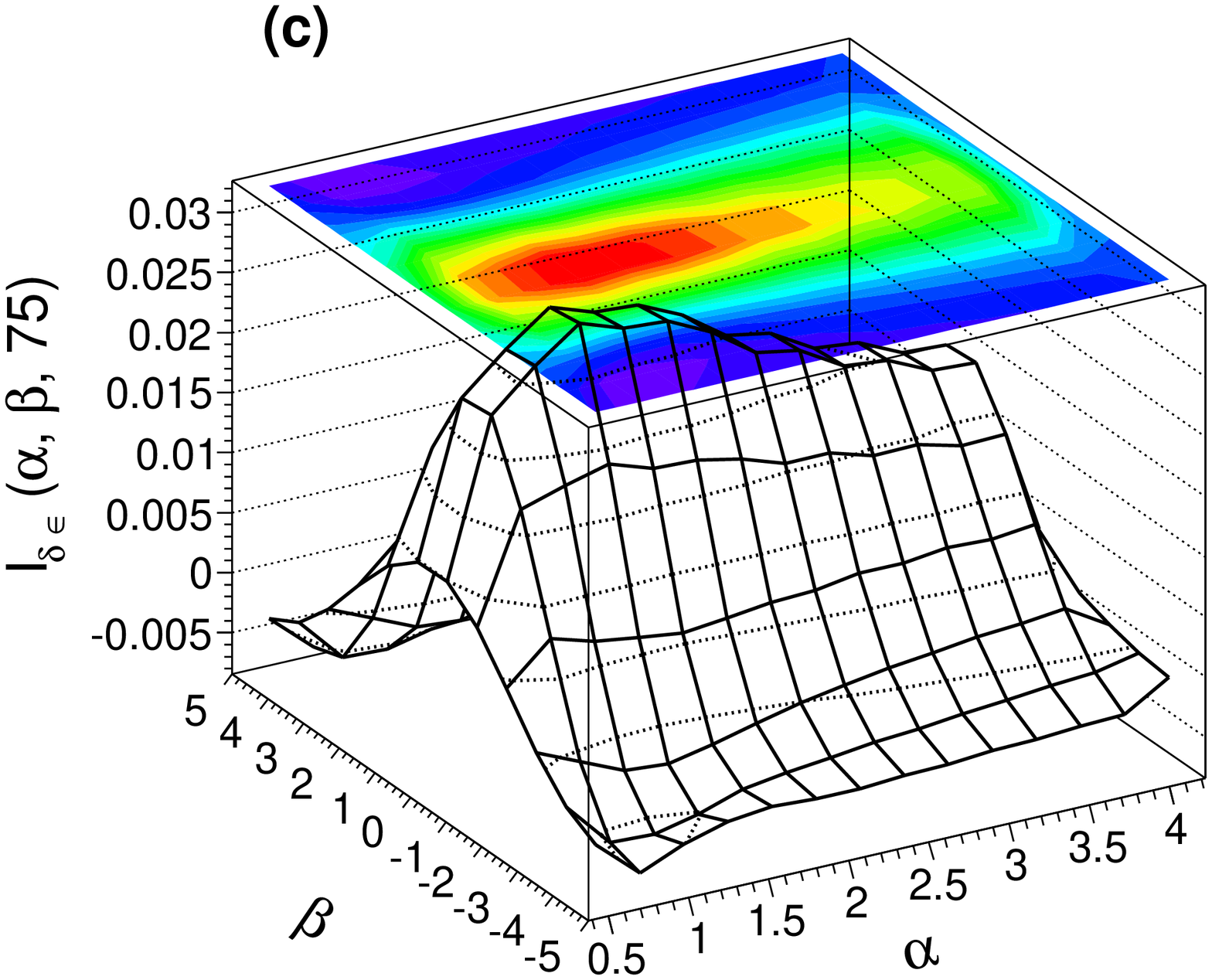}\includegraphics[scale=0.4]{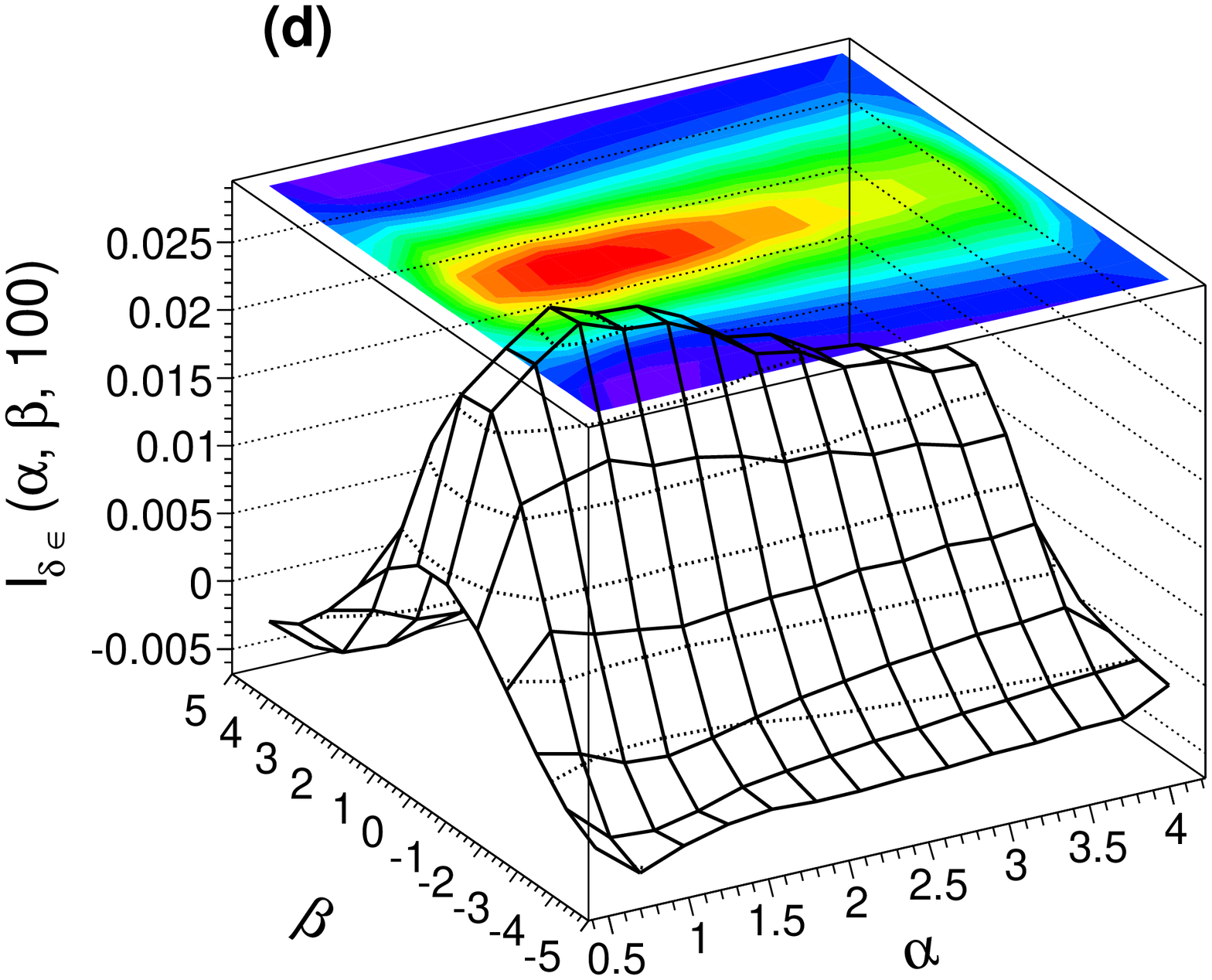}
}}
\caption{(Color online) Three dimensional plots (surfaces and contours) for
  $I_{g_{Tz}}$ and $I_{g_{\delta\epsilon}}$ as functions of $\alpha$,
  $\beta$ and $\kappa$ for $\eta/s=1/4\pi$. The plots are shown
  starting from a minimum value of $\alpha_{\mbox{\tiny{min}}}=0.5$
  and for values (left to right) of $\kappa= 75,\ 100$.}
\label{superficies}
\end{figure*}
When considering the effect of the derivative of the delta function in Eq.~(\ref{3}) for the integrations that lead to the energy and momentum components deposited into the medium, we can generically write
\bea
   \left. \int d\omega\ F(\omega)\frac{\partial}{\partial\omega}\delta (\omega - {\mathbf{k}}\cdot{\mathbf{v}}) =
   - \frac{\partial}{\partial\omega}F(\omega)\right|_{\omega = {\mathbf{k}}\cdot{\mathbf{v}}}.
\label{generical}
\eea
Also, the dependence on $\omega$ of this function is a product of the form $F(\omega)=e^{-i\omega t}f(\omega)$, thus
\bea
   \frac{\partial}{\partial\omega}F(\omega)=
   -ite^{-i\omega t}f(\omega) + e^{-i\omega t} \frac{\partial}{\partial\omega}f(\omega).
\label{form}
\eea
Therefore the total energy or momentum deposited into the medium can be expressed in terms of two contributions: The first term in Eq.~(\ref{form}) which corresponds to the one computed in Ref.~\cite{ADY} as if the energy per unit length was constant, multiplied by the time interval $t$, and the second one in that equation, which corresponds to a new contribution stemming from the  derivative of the function multiplying the exponential in the integrands. We write these generic contributions as
\bea
\mathbf{F}\left(\n{x},t\right)=vC\left[\n{F}_{0}\left(\n{x},t\right)t+\tilde{\n{F}}\left(\n{x},t\right)\right].
\label{4}
\eea

\begin{figure*}[t]
{\centering
{\includegraphics[scale=0.4]{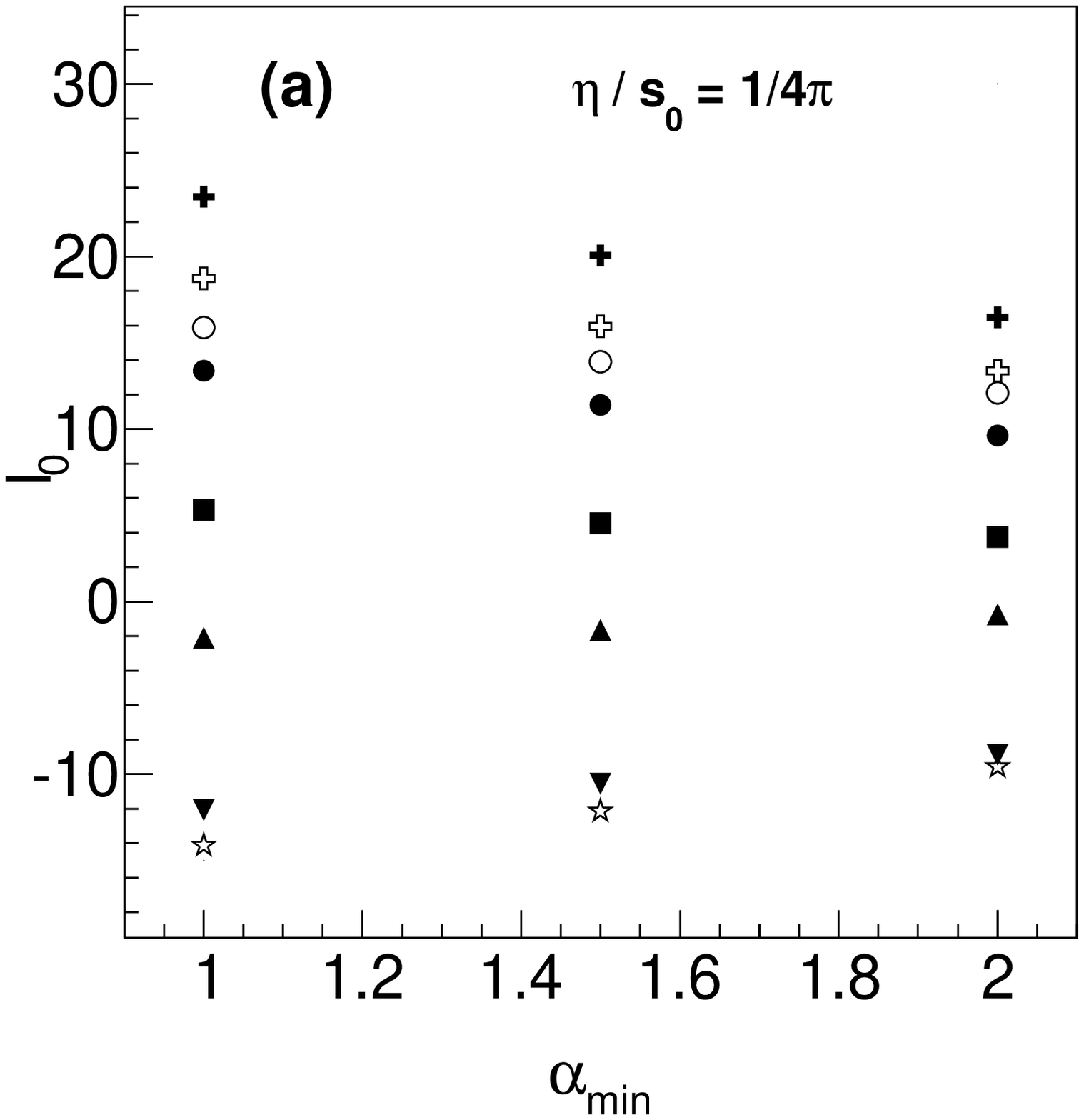}\includegraphics[scale=0.4]{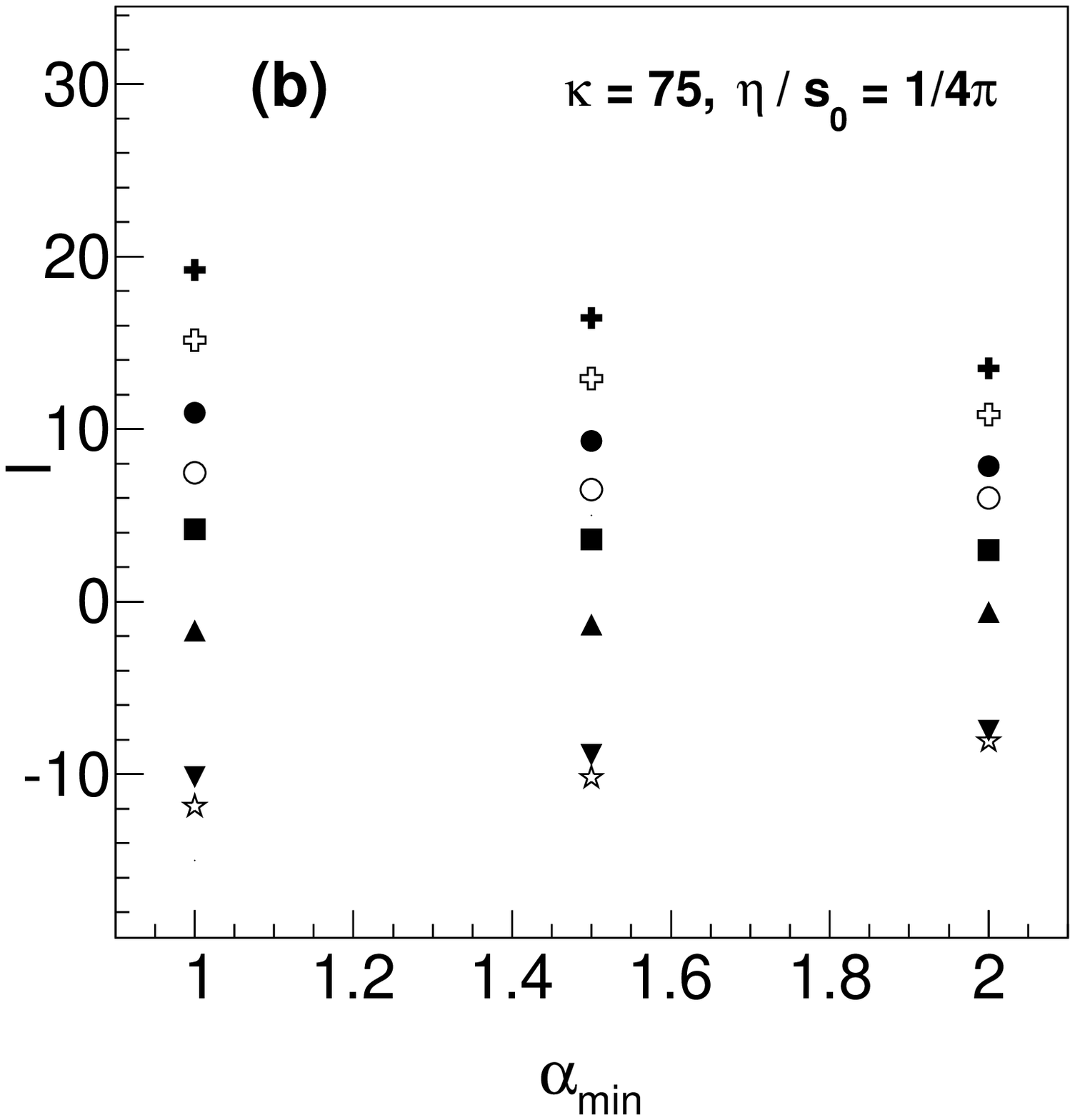}
\\\includegraphics[scale=0.4]{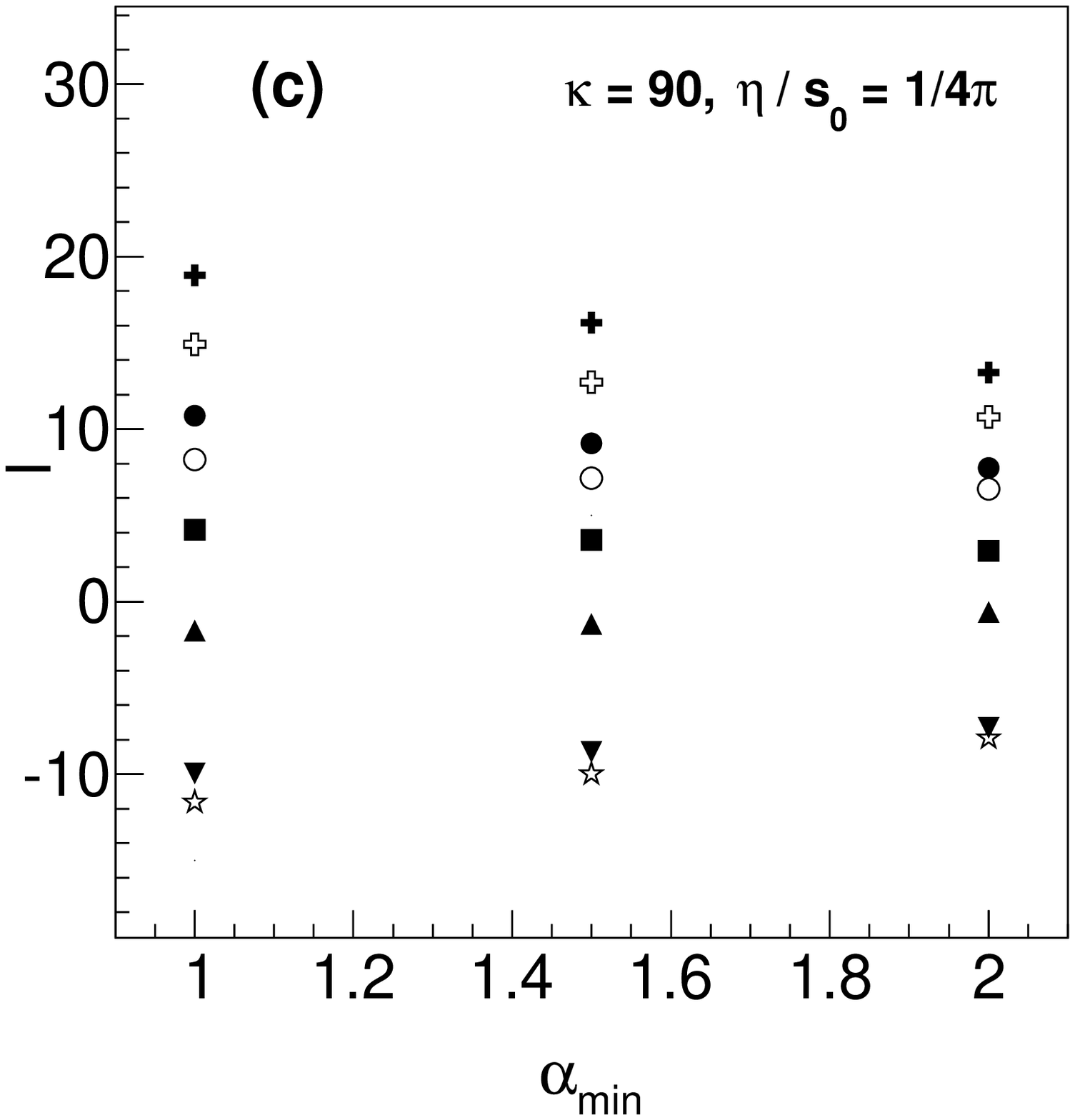}\includegraphics[scale=0.4]{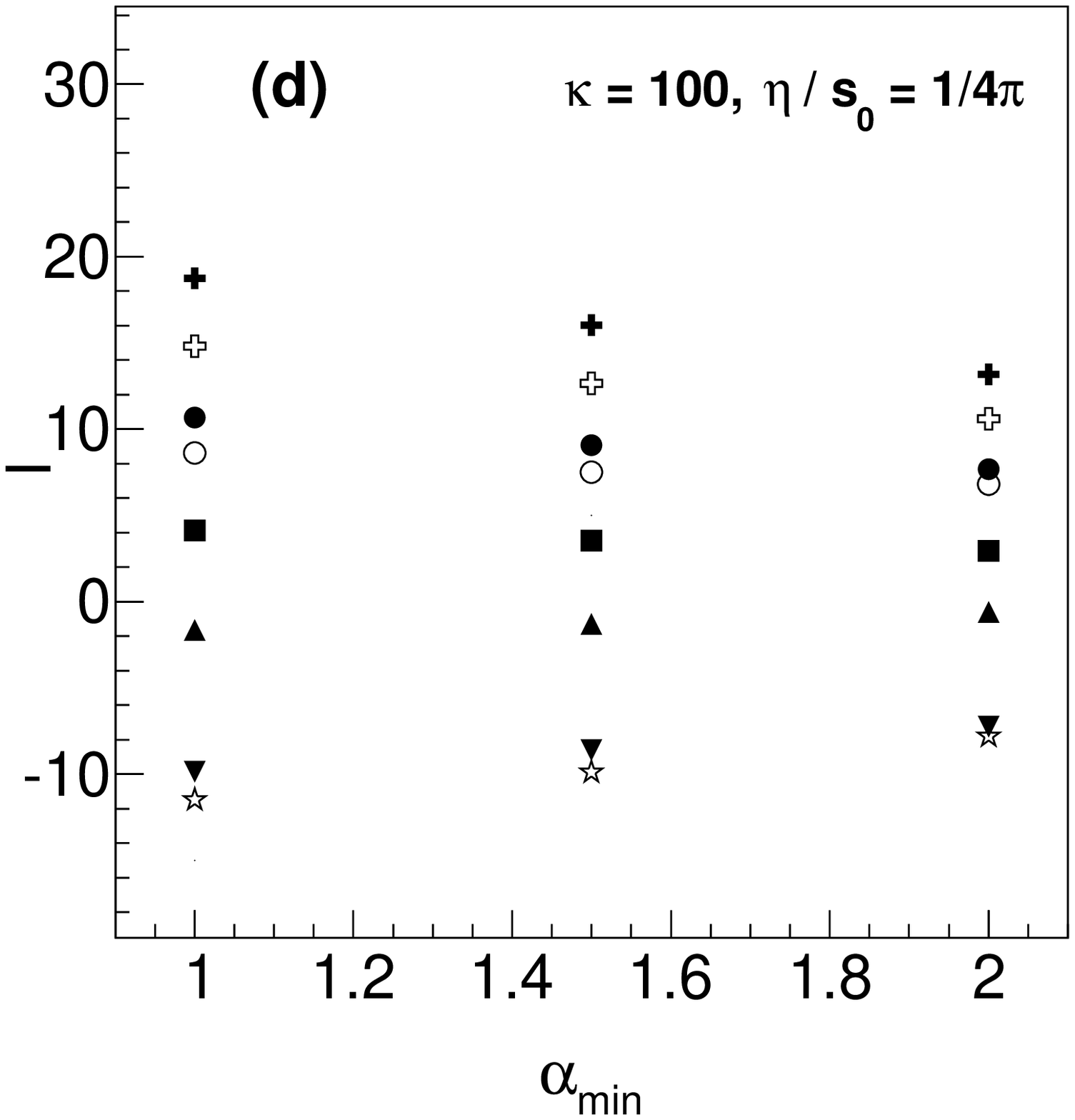}
\\\includegraphics[scale=0.5]{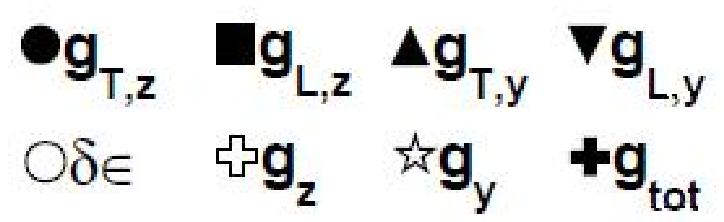}
}}
\caption{Integrals of the functions $I_{g_{Tz}}(\kappa), I_{g_{Ty}}(\kappa), I_{g_{Lz}}(\kappa), I_{g_{Ly}}(\kappa)$,
$I_{g_{\delta\epsilon}}(\kappa)$, $I_{g_{z}}(\kappa)$ and $I_{g_{y}}(\kappa)$, defined in Eqs.~(\ref{finalenergy})-(\ref{finalmomentum}), over the domain
$\alpha_{\mbox{\tiny{min}}}<\alpha<6$,
$-5<\beta<5$ for the different values of
$\alpha_{\mbox{\tiny{min}}}$ and $\kappa= 75,\ 90,\ 100$, with $\eta/s=1/4\pi$. Also the constant energy-loss $I_0$ case is plotted for comparison purposes. Notice that for all of values of
$\alpha_{\mbox{\tiny{min}}}$, the hierarchy of modes remains the same as for the case with constant $dE/dx$ and energy-momentum is preferentially deposited along
the head-shock.}
\label{I1}
\end{figure*}

In order to make the analysis more transparent, let us consider that $t$ represents a parameter that accounts for the time during which the parton travels trough the medium. For a hydrodynamical description, we require that this time is large enough compared to the sound attenuation length. Thus, it is convenient to express this time in units of $\Gamma_s$, introducing a dimensionless phenomenological quantity $\kappa$, given by
\bea
Ct=C_\kappa\frac{t}{\left(\frac{3\Gamma_s}{2v}\right)}\equiv C_\kappa\kappa,
\label{kappa}
\eea
where $\kappa=(3\Gamma_s /2v)^{-1}t$ is a characteristic time scale given in units of the sound attenuation length and $C_\kappa$ is a dimensionless free parameter that will be fixed by requiring that the total energy and momentum deposited within the medium by the fast moving parton is the same as in the case of a constant $dE/dx$. With this definition the energy and momentum deposited into the medium can be written as
\bea
\n{F}(\mathbf{x},t)&=&C_\kappa v\left[\kappa\mathbf{F}_0(\mathbf{x},t)+\left(\frac{2v}{3\Gamma_s}\right)\T{\n{F}}(\mathbf{x},t)\right].
\label{form2}
\eea
\begin{figure*}[t]
{\centering
{\includegraphics[scale=0.4]{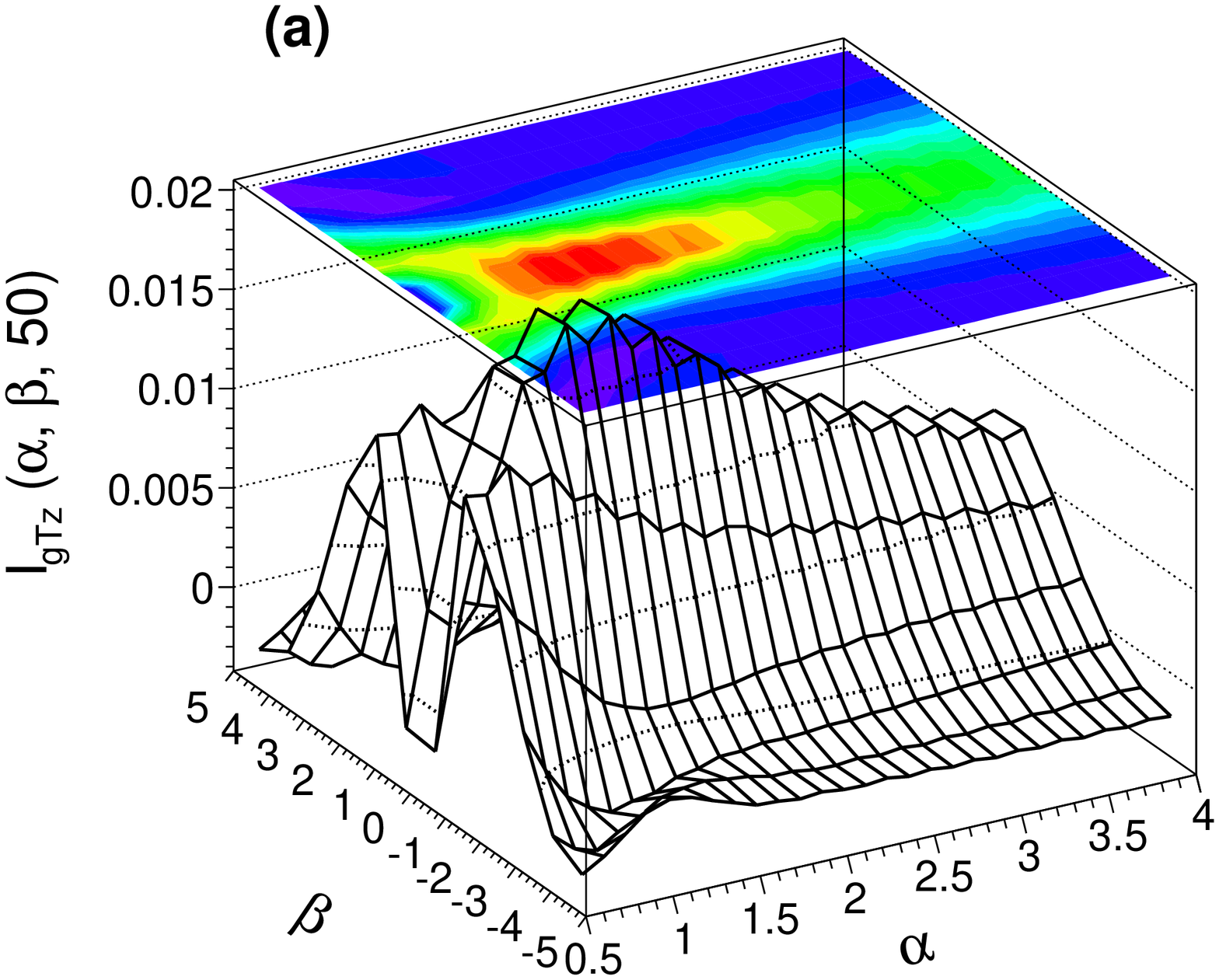}\includegraphics[scale=0.4]{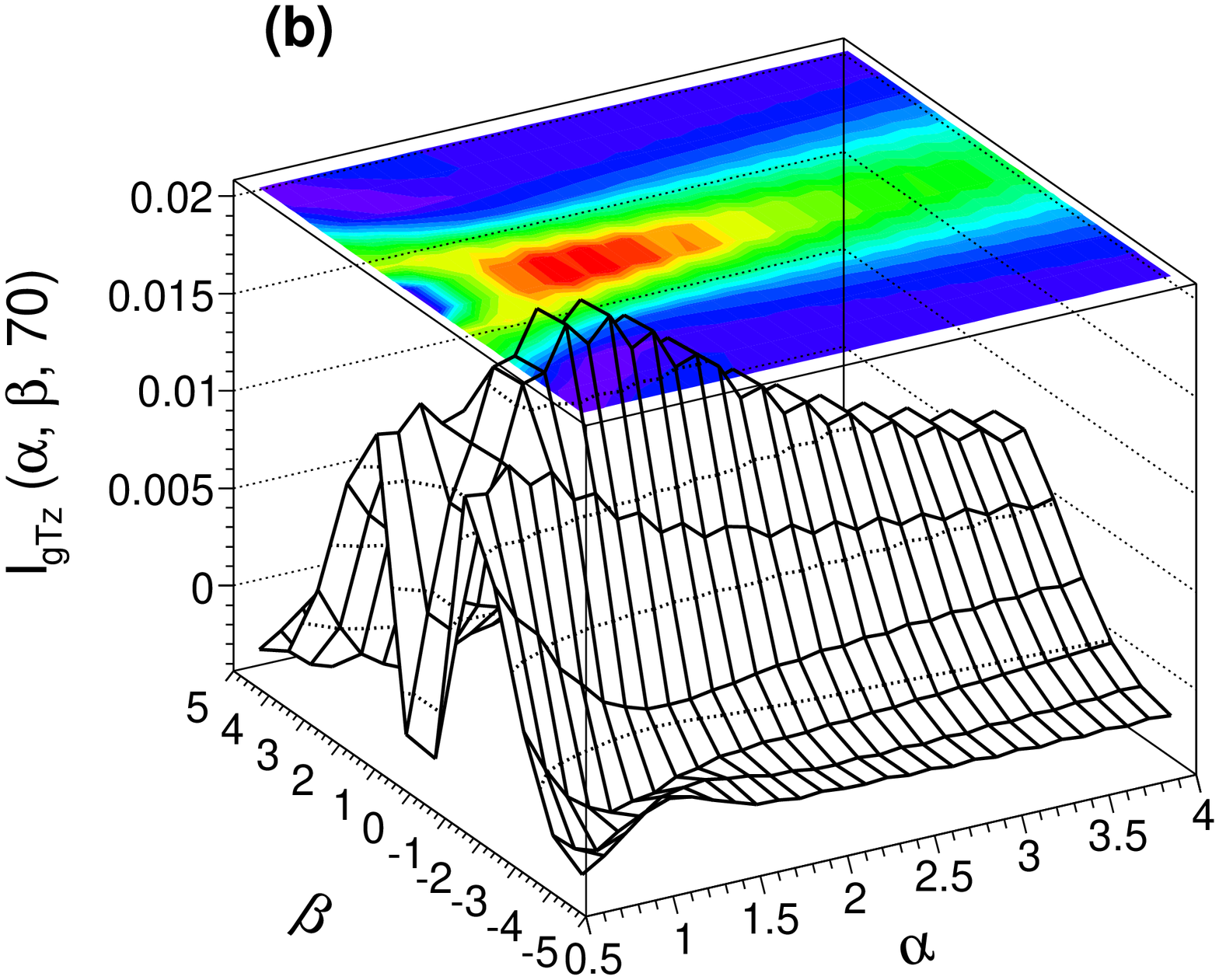}
\\\includegraphics[scale=0.4]{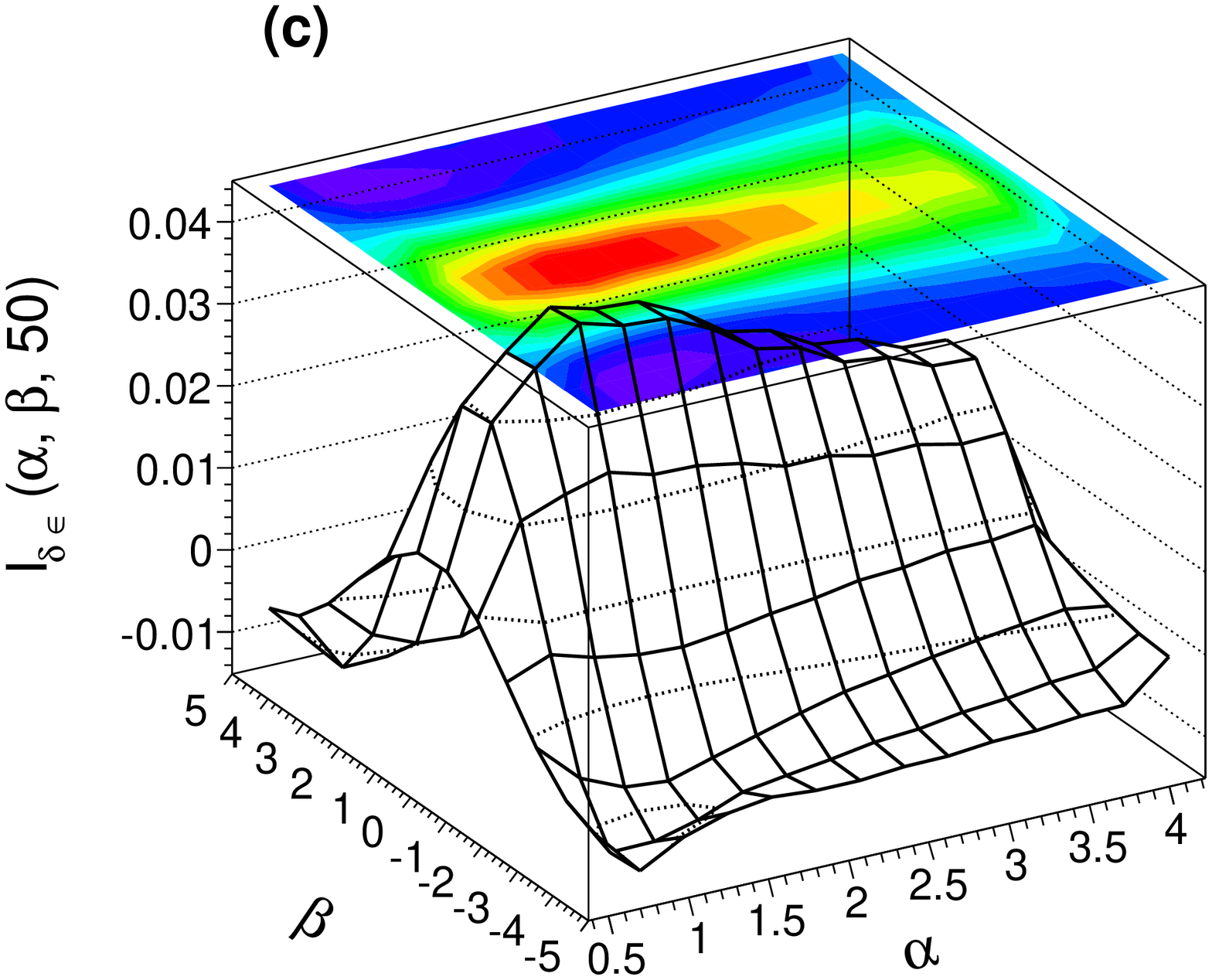}\includegraphics[scale=0.4]{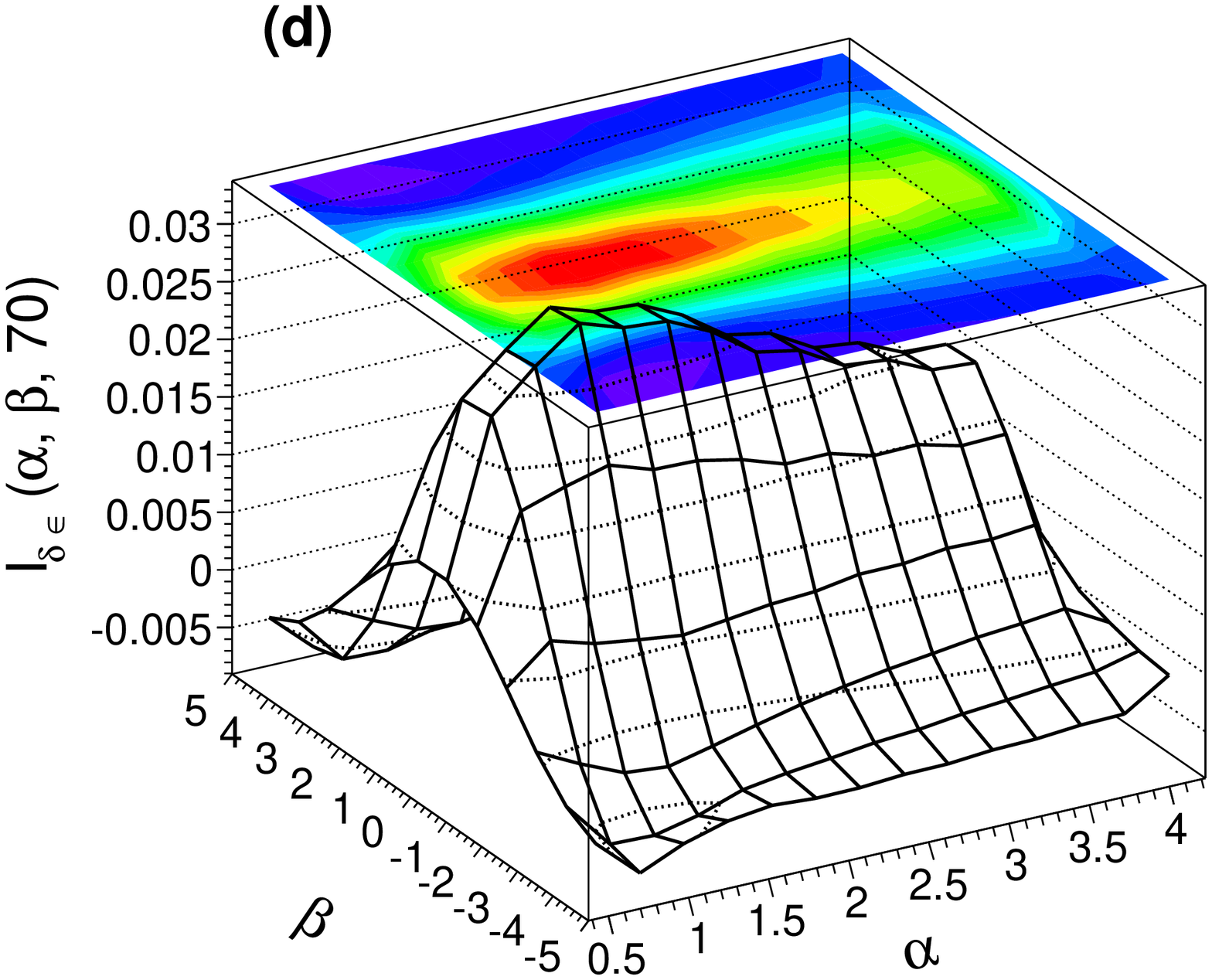}
}}
\caption{(Color online) Three dimensional plots (surfaces and contours) for $I_{g_{Tz}}$ and $I_{g_{\delta\epsilon}}$ as functions of $\alpha$, $\beta$ and $\kappa$ for $\eta/s=1.5/4\pi$. The plots are shown starting from a minimum value of $\alpha_{\mbox{\tiny{min}}}=0.5$ and for values (left to right) of $\kappa= 50,\ 70$.}
\label{superficies2}
\end{figure*}
We can now use Eqs.~(\ref{eps})--(\ref{gt}) to obtain the space-time solutions for $\delta\epsilon({\mathbf{x}},t)$ and ${\mathbf{g}}({\mathbf{x}},t)$. Using Eq.~(\ref{FT}) and after integration in $\omega$, the new contributions are
\bea
\tilde{\mathbf{g}}_{T}(\mathbf{x},t)&=&\int{\frac{d^3k}{(2\pi)^3}e^{i\mathbf{k}\cdot(\mathbf{x}-\mathbf{v}t)}}\nn
&\times&\left[\mathbf{v}-\frac{(\mathbf{k}\cdot\mathbf{J})\mathbf{k}}{k^{2}} \right]\frac{1}{(\mathbf{k}\cdot\mathbf{v}+i\frac{3}{4}\Gamma_{s} k^2)^2},
\label{gtnew}
\eea
\bea
\tilde{\mathbf{g}}_L(\mathbf{x},t)=-\tilde{\mathbf{g}}_{L1}(\mathbf{x},t)+\tilde{\mathbf{g}}_{L2}(\mathbf{x},t),
\label{glnew}
\eea
with
\begin{figure*}[t]
{\centering
{\includegraphics[scale=0.4]{fig2a.eps}\includegraphics[scale=0.4]{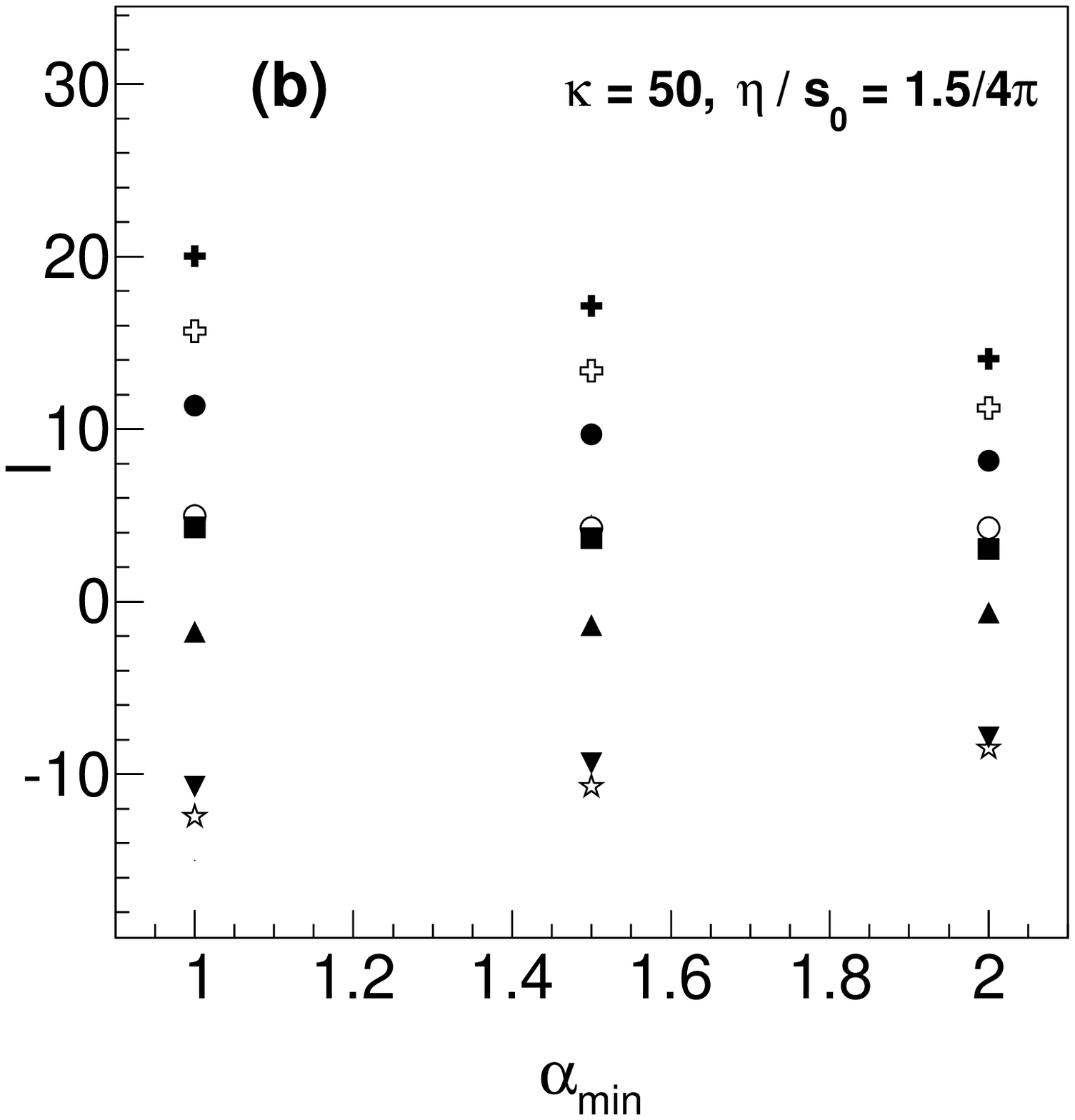}
\\\includegraphics[scale=0.4]{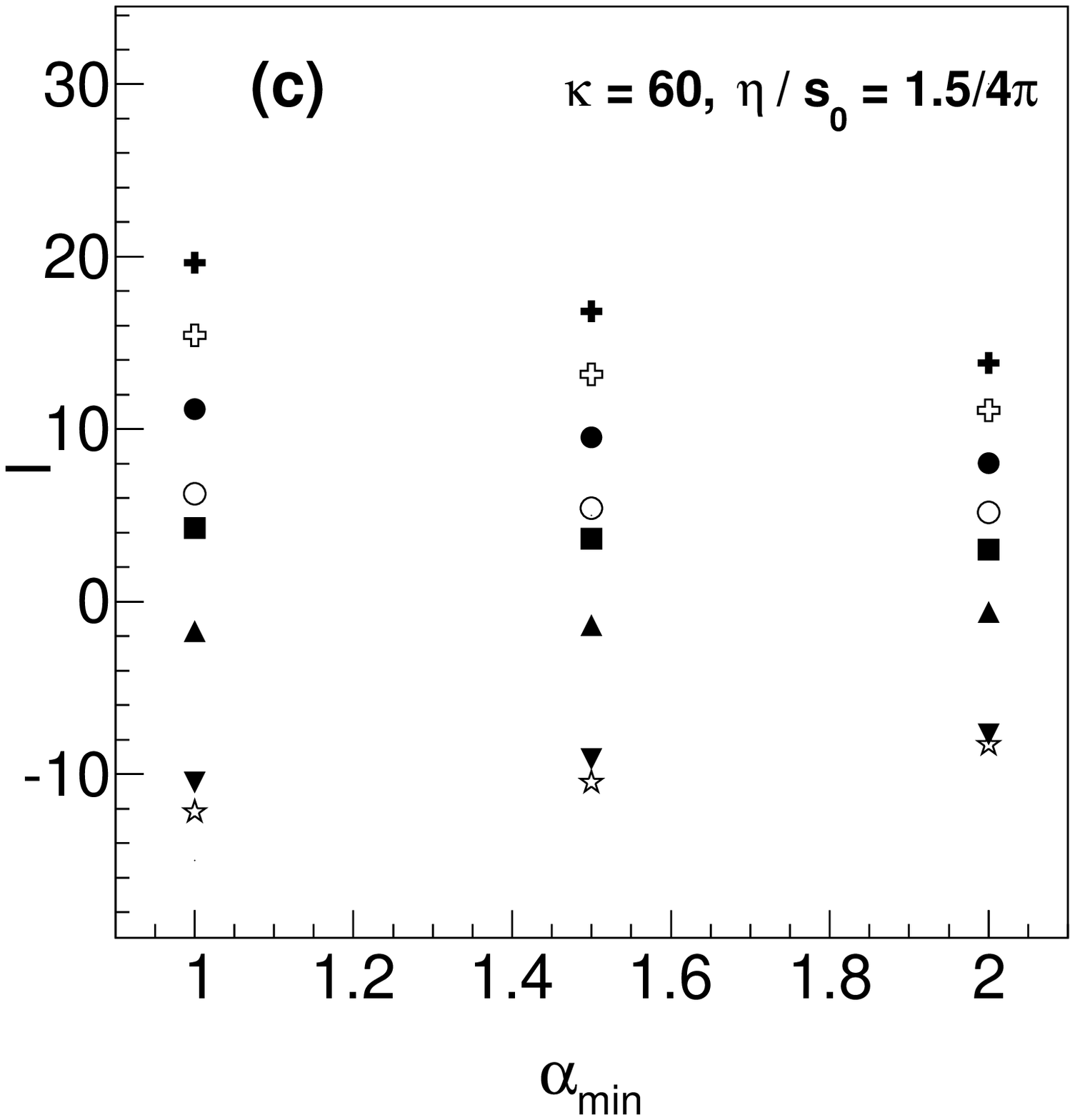}\includegraphics[scale=0.4]{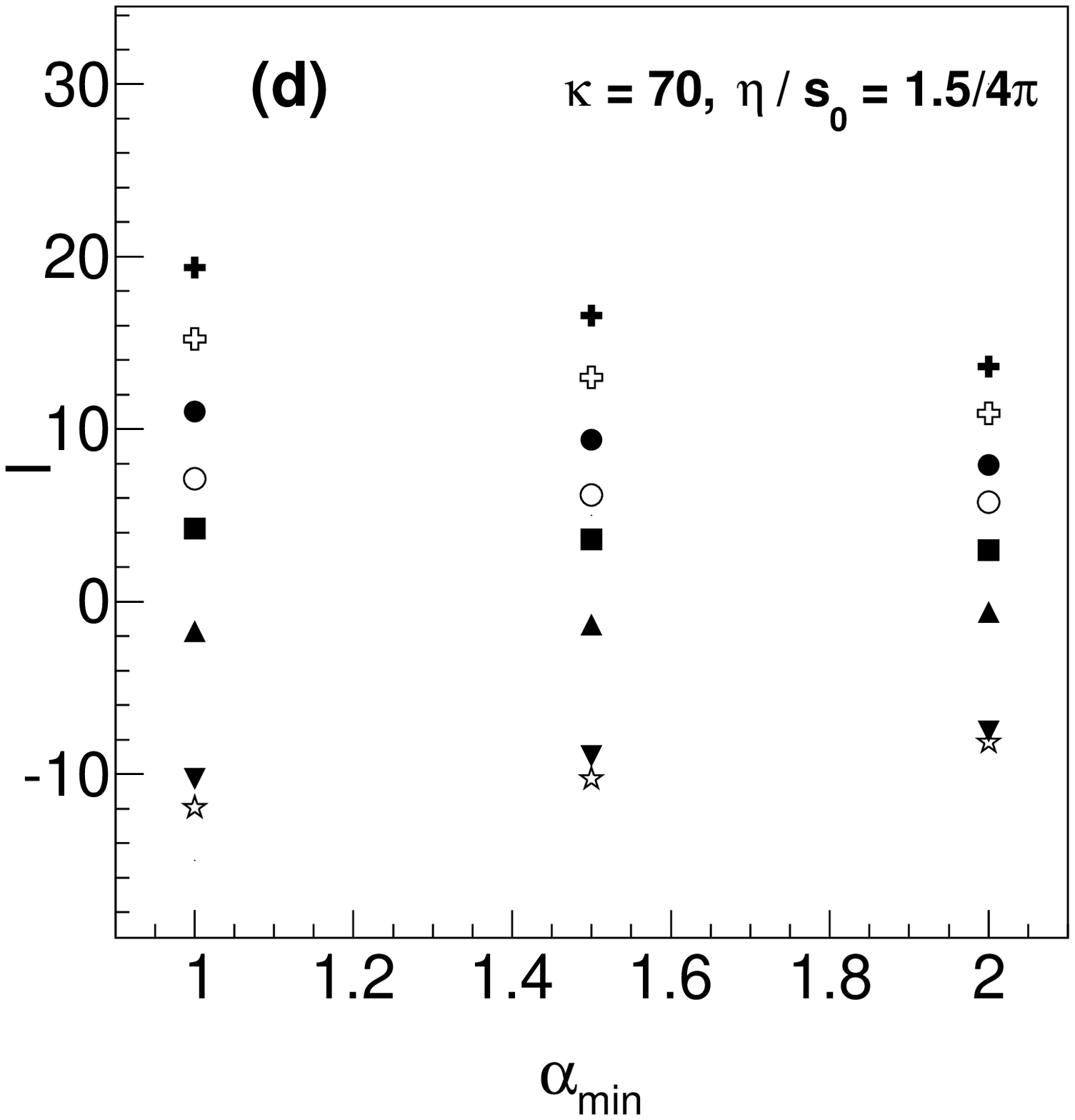}
\\\includegraphics[scale=0.5]{simbolos.eps}
}}
\caption{Integrals of the functions $I_{g_{Tz}}(\kappa), I_{g_{Ty}}(\kappa), I_{g_{Lz}}(\kappa), I_{g_{Ly}}(\kappa)$,
$I_{g_{\delta\epsilon}}(\kappa)$, $I_{g_{z}}(\kappa)$ and $I_{g_{y}}(\kappa)$, Eqs.~(\ref{finalenergy})-(\ref{finalmomentum}), over the domain
$\alpha_{\mbox{\tiny{min}}}<\alpha<6$,
$-5<\beta<5$ for the different values of
$\alpha_{\mbox{\tiny{min}}}$ and $\kappa= 50,\ 60,\ 70$, with $\eta/s=1.5/4\pi$. Also the constant energy-loss $I_0$ case with $\eta/s=1/4\pi$ is plotted for comparison purposes. Notice that for all of values of
$\alpha_{\mbox{\tiny{min}}}$, the hierarchy of modes remains  the same as for the case with constant $dE/dx$ and energy-momentum is preferentially deposited along
the head-shock.}
\label{I2}
\end{figure*}
\bea
\T{\n{\mathbf{g}}}_{L1(\mathbf{x},t)}&=&\int\frac{d^3k}{(2\pi)^3}e^{\mathbf{k}\cdot(\mathbf{x}-\mathbf{v}t)}\nn
&\times&\frac{(\mathbf{k}\cdot\mathbf{v})\mathbf{k}}{k^2\left[(\mathbf{k}\cdot\mathbf{v})^2-c_s^2k^2+i\Gamma_s(\mathbf{k}\cdot\mathbf{v}) k^2\right]},
\label{gl1}
\eea
\begin{eqnarray}
\T{\n{\mathbf{g}}}_{L2}(\mathbf{x},t)&=&\int\frac{d^3k}{(2\pi)^3}\mathbf{k}e^{\mathbf{k}\cdot(\mathbf{x}-\mathbf{v}t)}\nn
&\times&\frac{\left(\frac{(\mathbf{k}\cdot\mathbf{v})^2}{k^2}+c_s^2\right)\left(2\mathbf{k}\cdot\mathbf{v}+i\Gamma_sk^2\right)}{\left[(\mathbf{k}\cdot\mathbf{v})^2-c_s^2k^2+i\Gamma_s(\mathbf{k}\cdot\mathbf{v}) k^2\right]^2},
\label{gl2}
\end{eqnarray}
and
\bea
\delta\tilde{\epsilon}(\mathbf{x},t)=\delta\tilde{\epsilon}_1(\mathbf{x},t)-\delta\tilde{\epsilon}_2(\mathbf{x},t)
\label{de}
\eea
with
\begin{figure*}[t]
{\centering
{\includegraphics[scale=0.4]{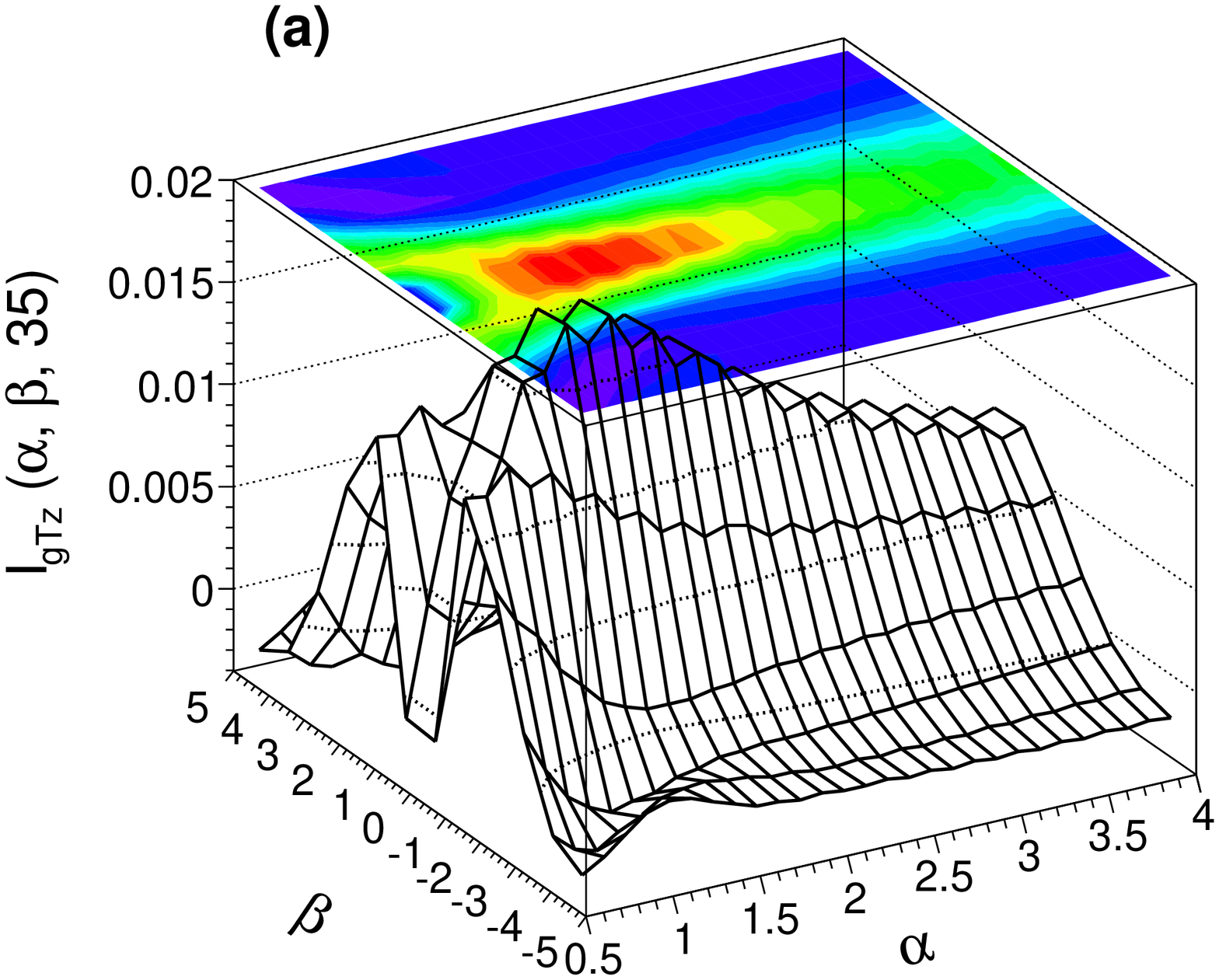}\includegraphics[scale=0.4]{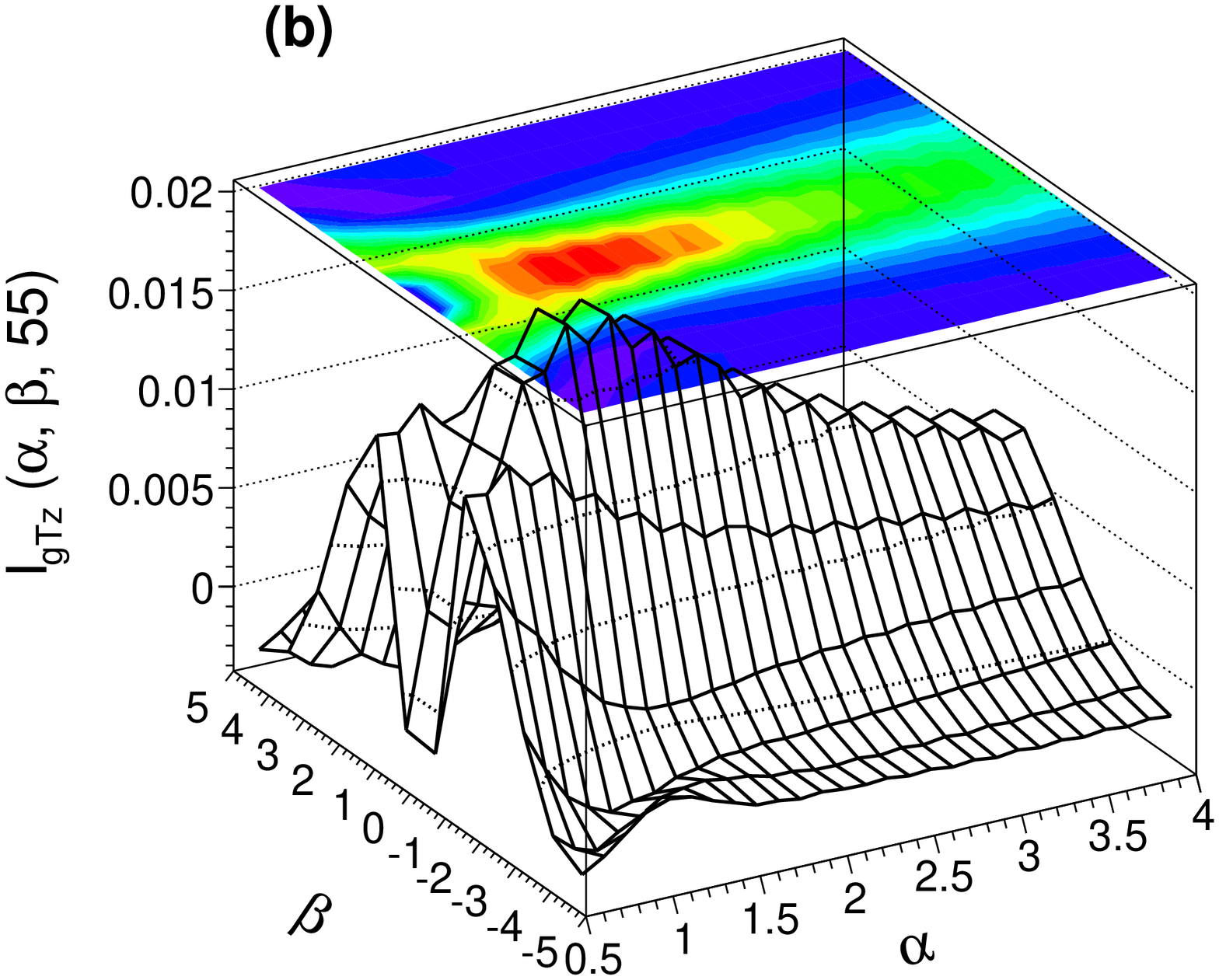}
\\\includegraphics[scale=0.4]{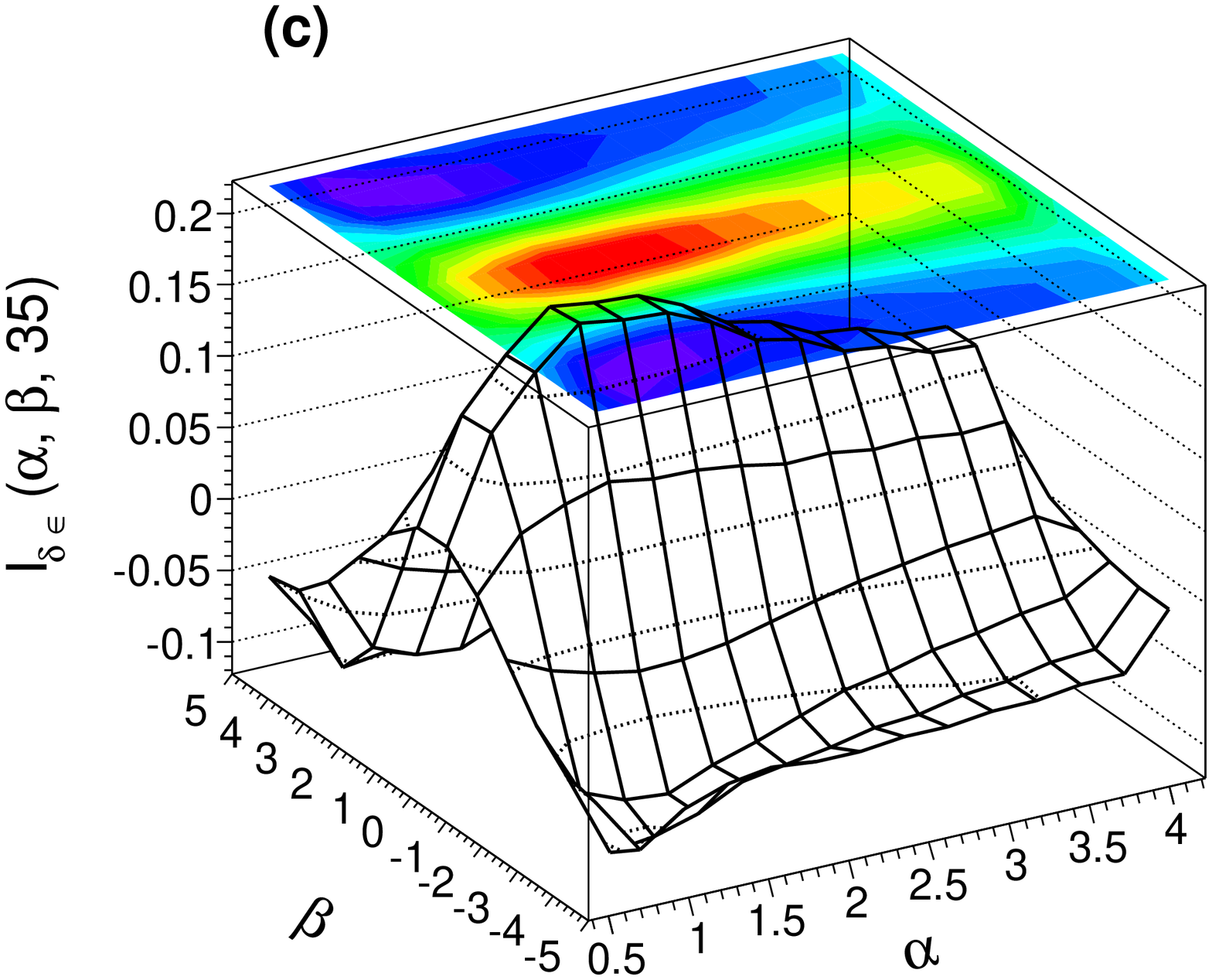}\includegraphics[scale=0.4]{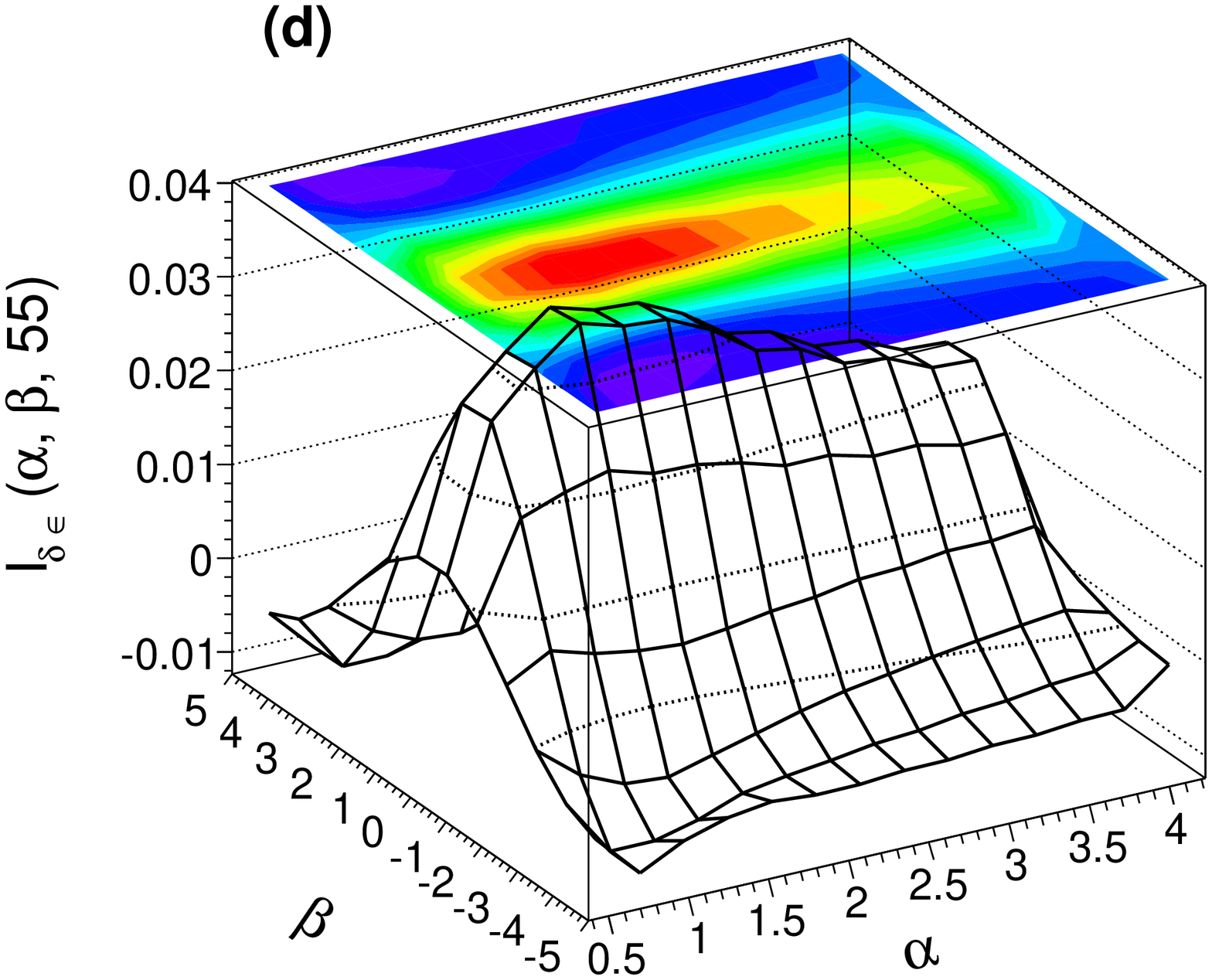}
}}
\caption{(Color online) Three dimensional plots (surfaces and contours) for $I_{g_{Tz}}$ and $I_{g_{\delta\epsilon}}$ as functions of $\alpha$, $\beta$ and $\kappa$ for $\eta/s=2/4\pi$. The plots are shown starting from a minimum value of $\alpha_{\mbox{\tiny{min}}}=0.5$ and for values (left to right) of $\kappa= 35,\ 55$.}
\label{superficies3}
\end{figure*}
\bea
\delta\tilde{\epsilon}_1(\mathbf{x},t)=-\int\frac{d^3k}{(2\pi)^3}\frac{e^{\mathbf{k}\cdot(\mathbf{x}-\mathbf{v}t)}}{(\mathbf{k}\cdot\mathbf{v})^2-c_s^2k^2+i\Gamma_s(\mathbf{k}\cdot\mathbf{v}) k^2},\nn
\label{de1}
\eea
\bea
\delta\tilde{\epsilon}_2(\mathbf{x},t)&=& i \int\frac{d^3k}{(2\pi)^3}e^{\mathbf{k}\cdot(\mathbf{x}-\mathbf{v}t)}\nn
&\times&\frac{\left(2i\mathbf{k}\cdot\mathbf{v}-\Gamma_sk^2\right)\left(2\mathbf{k}\cdot\mathbf{v}+i\Gamma_sk^2\right)}{\left[(\mathbf{k}\cdot\mathbf{v})^2-c_s^2k^2+i\Gamma_s(\mathbf{k}\cdot\mathbf{v}) k^2\right]^2}.\nn
&&
\label{de2}
\eea
In order to compute the integrals in Eqs.~(\ref{gtnew})--(\ref{de2}) we use cylindrical coordinates with $k_z$ directed along the direction of motion ${\mathbf{v}}$ of the fast parton. Let us look in detail at the computation of the $z$-component of $\T{g}_T$. After carrying out the angular integration we get
\begin{figure*}[t]
{\centering
{\includegraphics[scale=0.4]{fig2a.eps}\includegraphics[scale=0.4]{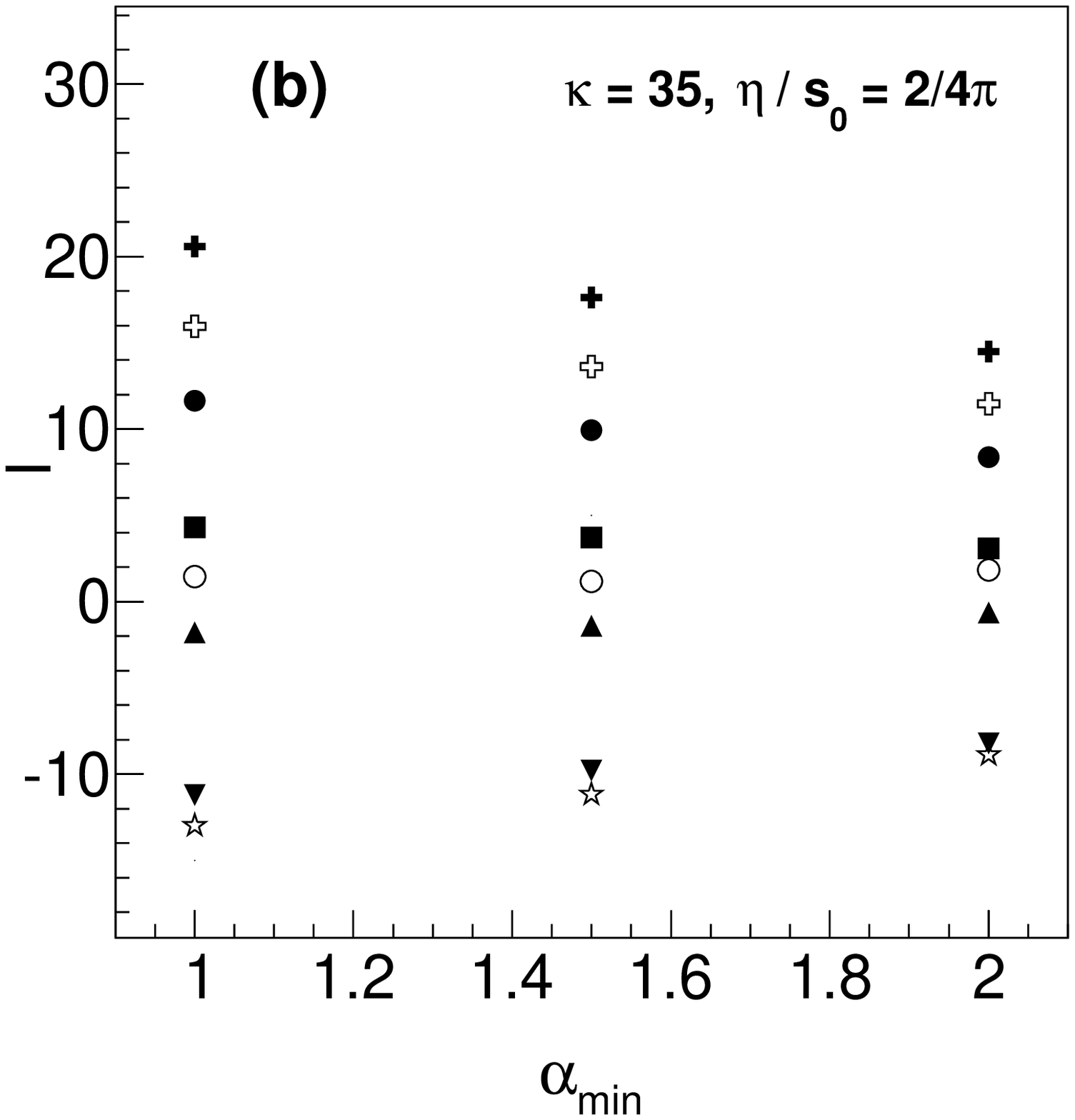}
\\\includegraphics[scale=0.4]{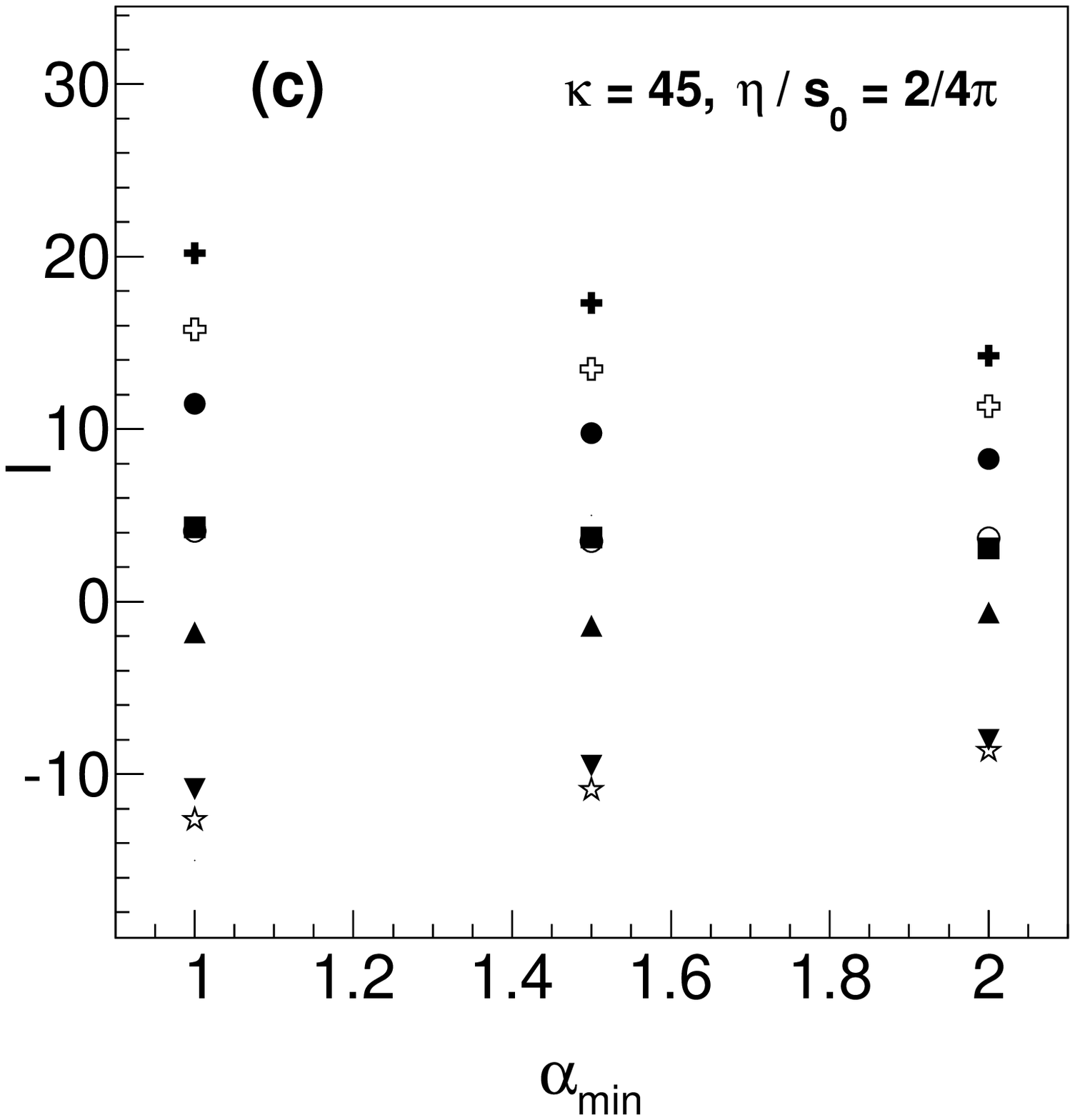}\includegraphics[scale=0.4]{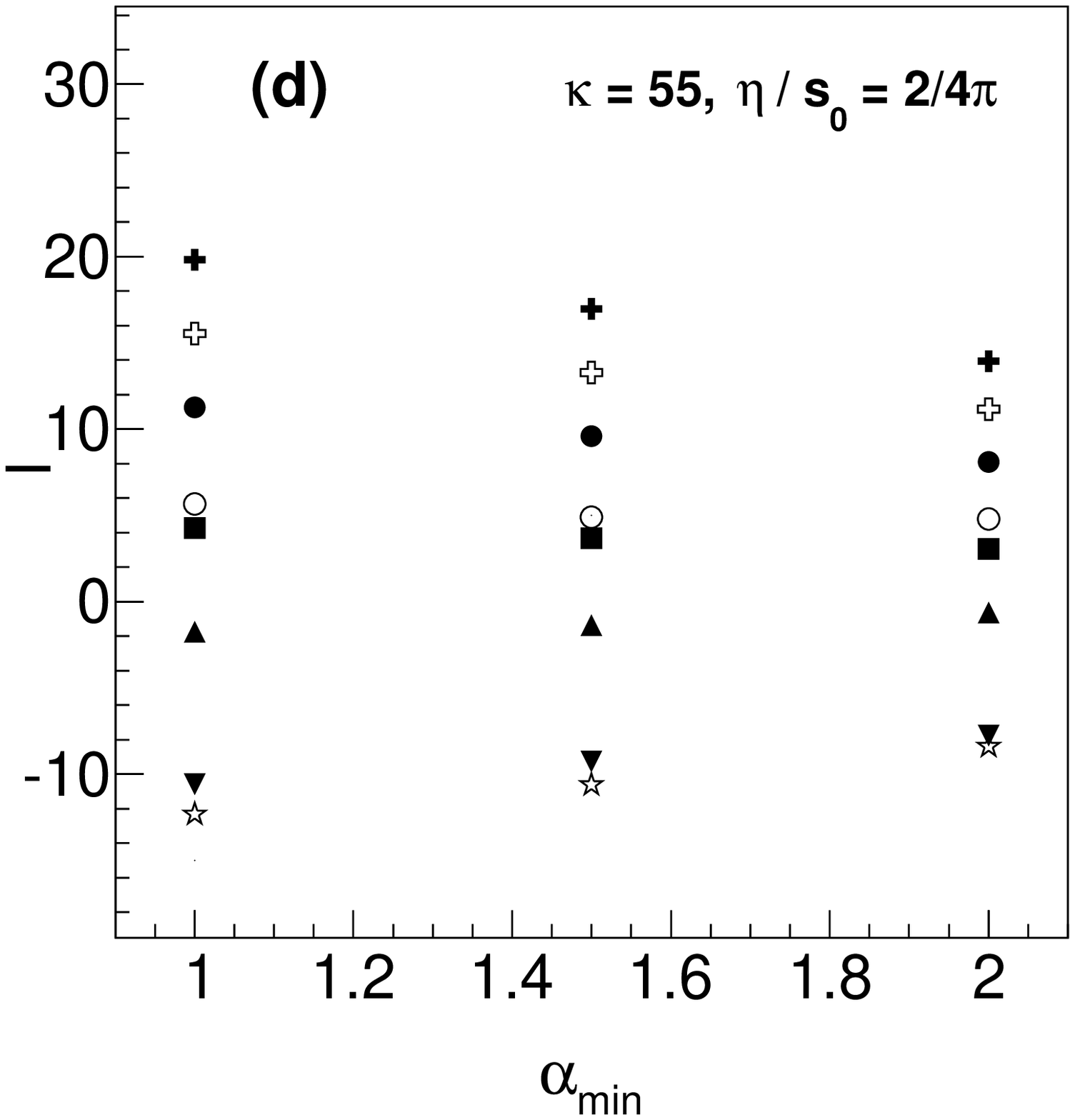}
\\\includegraphics[scale=0.5]{simbolos.eps}
}}
\caption{Integrals of the functions $I_{g_{Tz}}(\kappa), I_{g_{Ty}}(\kappa), I_{g_{Lz}}(\kappa), I_{g_{Ly}}(\kappa)$,
$I_{g_{\delta\epsilon}}(\kappa)$, $I_{g_{z}}(\kappa)$ and $I_{g_{y}}(\kappa)$, defined in Eqs.~(\ref{finalenergy})-(\ref{finalmomentum}), over the domain
$\alpha_{\mbox{\tiny{min}}}<\alpha<6$,
$-5<\beta<5$ for the different values of
$\alpha_{\mbox{\tiny{min}}}$ and $\kappa= 35,\ 45,\ 55$, with $\eta/s=2/4\pi$. Also the constant energy-loss $I_0$ with $\eta/s=1/4\pi$ is plotted for comparison purposes. Notice that for all of values of
$\alpha_{\mbox{\tiny{min}}}$ the hierarchy of modes remains the same as for the case with constant $dE/dx$ and energy-momentum is preferentially deposited along
the head-shock.}
\label{I3}
\end{figure*}
\bea
(\T{g}_{T})_z&=& v \int_{0}^{\infty}\frac{dk_{T}}{(2\pi)^2}\int_{-\infty}^{\infty}dk_ze^{ik_z(z-vt)}\nonumber\\
&\times&\frac{k_T^3{\mathcal{J}_0}(k_Tx_T)}{(k_zv+i\frac{3}{4}\Gamma_{s} k^2)^2(k_T^2+k_z^2)}\nonumber\\
&=&2\pi i v \int_{0}^{\infty}\frac{dk_{T}}{(2\pi)^2}k_T^3{\mathcal{J}_0}(k_Tx_T)\nonumber\\
&\times&\left[\text{Res}1+\text{Res}2\right],
\label{gtz-z}
\eea
where ${\mathcal{J}_0}$ is a Bessel function and $x_T=\sqrt{y^2}$ is the distance
from the parton along the transverse direction (directed along the $\hat{y}$ axis,
in the geometry we are using) and Res1 and Res2 represent the residues at the two poles in the integrand of Eq.~(\ref{gtz-z}). To carry out the contour integration we close the contour on the lower half
$k_z$-plane in order to ensure causal motion ($z-vt <0$). The first residue is given by
\bea
\text{Res}1=-i\frac{e^{k_T(z-vt)}}{2k_T^3v^2},
\eea
which can be  analytically integrated with respect to $k_T$. For the second residue, we express the integral in terms of the dimensionless quantities
\bea
&&\xi\equiv\left(\frac{3\Gamma_s}{2v}\right)k_T, \,\,\,\alpha\equiv\left(\frac{3\Gamma_s}{2v}\right)^{-1}\left | z-vt \right |\nn
&&\beta\equiv\left(\frac{3\Gamma_s}{2v}\right)^{-1}x_T, \,\,\,s=\sqrt{1+\xi^2}-1,
\label{dimensionless}
\eea
thus
\begin{eqnarray}
\text{Res}2=\frac{i}{2v^2}\left(\frac{3\Gamma_s}{2v}\right)^{3}e^{-\alpha s}\frac{s-\alpha (s+1)}{s(s+1)^3}.
\label{res2}
\end{eqnarray}
Note that the variables $\alpha$ and $\beta$ represent the distance from the source in the parton direction of motion and in the transverse direction, respectively, in units of the sound attenuation length, whereas $\xi$ is the transverse momentum in units of the inverse of the sound attenuation length.

Putting all together, the integral in Eq~(\ref{gtz-z}) becomes
\begin{eqnarray}
(\T{g}_{T})_z&=&\frac{1}{v}\left(\frac{1}{4\pi}\right)\left(\frac{2v}{3\Gamma_s}\right)\Big[\frac{1}{\sqrt{\alpha^2+\beta^2}}\nonumber\\
&-&\int_{0}^{\infty}ds\frac{s-\alpha (s+1)}{(s+1)^2}(s+2)\nonumber\\
&\times&{\mathcal{J}_0}\left(\beta\sqrt{s(s+2)}\right)e^{-\alpha s}\Big]\nn
&\equiv&\left(\frac{1}{4\pi}\right)\left(\frac{2v}{3\Gamma_s}\right)\T{I}_{g_{Tz}}.
\label{gtzs}
\end{eqnarray}
In a similar fashion we get

\begin{eqnarray}
(\T{g}_T)_y&=&\frac{1}{v}\left(\frac{1}{4\pi}\right)\left(\frac{2v}{3\Gamma_s}\right)\left[\frac{\alpha-\sqrt{\alpha^2+\beta^2}}{\beta\sqrt{\alpha^2+\beta^2}}\right.\nn
&+&\left.\int_{0}^{\infty}ds{\mathcal{J}_1}(\beta\sqrt{s(s+2)})e^{-\alpha s}\right.\nn
&\times&\left.\sqrt{s(s+2)}\frac{(s^2+s+1)-\alpha s(s+1)}{(s+1)^2}\right]\nn
&\equiv&\left(\frac{1}{4\pi}\right)\left(\frac{2v}{3\Gamma_s}\right)\T{I}_{g_{Ty}},
\label{gtys}
\end{eqnarray}
where ${\mathcal{J}_1}$ is a Bessel function. For the $(\T{g}_{L1})_z$ component, after carrying out the angular integration we get
\begin{eqnarray}
(\T{g}_{L1})_z&=&\frac{1}{v}\int\frac{dk_T}{(2\pi)^2}\int dk_z k_z^2k_T{\mathcal{J}_0}(x_Tk_T)e^{ik_z(z-vt)}\nn
&\times&\frac{1}{(k_T^2+k_z^2)\left[k_z^2+\left(-\frac{c_s^2}{v^2}+i\Gamma_s\frac{k_z}{v}\right)(k_T^2+k_z^2)\right]}.\nn
\label{glz1}
\end{eqnarray}

For conditions close to the ones after a heavy-ion reaction, the quantity $c_s^2/v^2$ is small, since for a fast moving (massless) parton $v\simeq 1$ and for a relativistic gas, $c_s \simeq \sqrt{1/3}$. Therefore, we can expand the integrand in Eq.~(\ref{glz1}) in this parameter. To first order in $c_s^2/v^2$, we get

\begin{eqnarray}
(\T{g}_{L1})_z&=&\frac{1}{v}\int\frac{dk_T}{(2\pi)^2}\int dk_z\frac{k_Tk_z^2{\mathcal{J}_0}(x_Tk_T)e^{ik_z(z-vt)}}{(k_T^2+k_z^2)}\nn
&\times&\left\{\frac{1}{k_z^2+i\Gamma_s\frac{k_z}{v}(k_T^2+k_z^2)}\right.\nn
&+&\left.\frac{(k_T^2+k_z^2)}{\left[k_z^2+i\Gamma_s\frac{k_z}{v}(k_T^2+k_z^2)\right]^2}\left(\frac{c_s^2}{v^2}\right)\right\}.
\label{glz1primerorden}
\end{eqnarray}
To perform the integral it is convenient to introduce the variable $r$ related to $\xi$ by $r=\frac{4}{3}\xi$. Once again, in order to describe causal motion ($z-vt <0$), we close the contour on the lower half  $k_z$-plane. The remaining integral is obtained after the change of variable $s=\sqrt{1+r^2}-1$ and given by
\begin{eqnarray}
(\T{g}_{L1})_z&=&\frac{1}{v}\left(\frac{1}{4\pi}\right)\left(\frac{2v}{3\Gamma_s}\right)\left\{-\frac{1}{\sqrt{\alpha^2+\beta^2}}\right.\nn
&+&\left.\int_{0}^{\infty} ds {\mathcal{J}_0}\left(\frac{3}{4}\beta\sqrt{s(s+2)}\right)e^{-\frac{3}{4}\alpha s}\right.\nn
&\times&\left.\left[\frac{3}{8}+\frac{3}{2}\left(\frac{c_s}{v}\right)^2(s+1)\left(1+\frac{3}{4}\alpha (s+1)\right)\right]\right\}\nonumber\\
&\equiv&\left(\frac{1}{4\pi}\right)\left(\frac{2v}{3\Gamma_s}\right)\T{I}_{g_{L1z}}
\label{glz1s}.
\end{eqnarray}

In a similar fashion we obtain

\begin{eqnarray}
(\T{g}_{L2})_z&=&\frac{1}{v}\left(\frac{1}{4\pi}\right)\left(\frac{2v}{3\Gamma_s}\right)\left\{-\frac{2}{\sqrt{\alpha^2+\beta^2}}\right.\nn
&+&\left.\frac{3}{4}\int_{0}^{\infty}ds\frac{{\mathcal{J}_0}\left(\frac{3}{4}\beta\sqrt{s(s+2)}\right)}{(s+1)^2}e^{-\frac{3}{4}\alpha s}\right.\nn
&\times&\left.\left[\frac{(2s^2+4s+3)-\frac{3}{2}\alpha s(s+1)}{2}\right.\right.\nn
&-&\left.\left. \frac{2(s^2+s+1)-\frac{3}{2}\alpha s(s+1)}{s}\left(\frac{c_s}{v}\right)^2\right.\right.\nn
&-&\left.\left.\frac{\frac{9}{4}\alpha^2s(s+1)-3\alpha(2s^2+3s+4)}{2(s+1)}\left(\frac{c_s}{v}\right)^2\right.\right.\nn
&-&\left.\left.\frac{2s^2+4s+5}{(s+1)^2}\left(\frac{c_s}{v}\right)^2\right]\right\}\nn
&\equiv&\left(\frac{1}{4\pi}\right)\left(\frac{2v}{3\Gamma_s}\right)\T{I}_{g_{L2z}},
\label{glz2s}
\end{eqnarray}

\begin{eqnarray}
(\T{g}_{L1})_y&=&\frac{1}{v}\left(\frac{1}{4\pi}\right)\left(\frac{2v}{3\Gamma_s}\right)\left\{\frac{\sqrt{\alpha^2+\beta^2}-\alpha}{\beta\sqrt{\alpha^2+\beta^2}}\right.\nn
&-&\left.\frac{3}{4}\int_{0}^{\infty}ds\frac{\sqrt{s(s+2)}}{s}e^{-\frac{3}{4}\alpha s}{\mathcal{J}_1}\left(\frac{3}{4}\beta\sqrt{s(s+2)}\right)\right.\nn
&\times&\left[\left.1+\frac{\frac{3}{2}\alpha s(s+1)+2(2s+1)}{(s+1)^2}\left(\frac{c_s}{v}\right)^2\right]\right\}\nn
&\equiv&\left(\frac{1}{4\pi}\right)\left(\frac{2v}{3\Gamma_s}\right)\T{I}_{g_{L1y}},
\label{gly1s}
\end{eqnarray}

\bea
(\T{g}_{L2})_y&=&\frac{1}{v}\left(\frac{1}{4\pi}\right)\left(\frac{2v}{3\Gamma_s}\right)\left\{2\frac{\alpha-\sqrt{\alpha^2+\beta^2}}{\beta\sqrt{\alpha^2+\beta^2}}\right.\nn
&+&\left.\frac{3}{4}\int_{0}^{\infty}ds\frac{\sqrt{s(s+2)}}{s(s+1)^2}e^{-\frac{3}{4}\alpha s}{\mathcal{J}_1}\left(\frac{3}{4}\beta\sqrt{s(s+2)}\right)\right.\nn
&\times&\left.\left[\frac{\frac{3}{2}\alpha s(s+1)-2(2s^2+3s+2)}{2}\right.\right.\nn
&+&\left.\left.\frac{2s-\frac{3}{2}\alpha (s+1)}{s+1}\left(\frac{c_s}{v}\right)^2+\frac{\frac{9}{4}\alpha^2 s^2}{(s+1)}\left(\frac{c_s}{v}\right)^2\right.\right.\nn
&-&\left.\left.\frac{\frac{2}{3}\alpha (2s^3+3s^2+3s+2)}{(s+1)^3}\left(\frac{c_s}{v}\right)^2\right.\right.\nn
&-&\left.\left.\frac{2(4s^3+7s^2+8s+2)}{(s+1)^3}\left(\frac{c_s}{v}\right)^2\right]\right\}\nn
&\equiv&\left(\frac{1}{4\pi}\right)\left(\frac{2v}{3\Gamma_s}\right)\T{I}_{g_{L2y}},
\label{gly2}
\eea

\bea
\delta\T{\epsilon}_1(\mathbf{x},t)&=&\frac{1}{v^2}\left(\frac{1}{4\pi}\right)\left(\frac{2v}{3\Gamma_s}\right)\frac{3}{2}\int_{0}^{\infty}ds\frac{e^{-\frac{3}{4}\alpha s}}{s}\nn
&\times& {\mathcal{J}_0}\left(\frac{3}{4}\beta\sqrt{s(s+2)}\right)\nn
&\times&\left[1+c_s^2\frac{\frac{3}{2}\alpha s(s+1)+2(s^2+4s+2)}{2s(s+1)^2}\right]\nn
&\equiv&\left(\frac{1}{4\pi}\right)\left(\frac{2v}{3\Gamma_s}\right)\T{I}_{\delta\epsilon 1}
\label{des}
\eea
and
\bea
\delta\T{\epsilon}_2(\mathbf{x},t)&=&\frac{1}{v^2}\left(\frac{1}{4\pi}\right)\left(\frac{2v}{3\Gamma_s}\right)\frac{3}{4}\int_{0}^{\infty}dse^{-\frac{3}{4}\alpha s}\nn
&\times&{\mathcal{J}_0}\left(\frac{3}{4}\beta\sqrt{s(s+2)}\right)\left[\frac{2(2s^2+3s+2)}{s(s+1)^2}\right.\nn
&-&c_s^2\left.\frac{\frac{9}{4}\alpha^2 s^2(s+1)^2-3\alpha s(2s^3+3s^2+3s+2)}{s^2(s+1)^4}\right.\nn
&-&\left.\frac{4(2s^4+12s^3+19s^2+16s+4)}{s^2(s+1)^4}c_s^2\right.\nn
&-&\left.\frac{\frac{3}{2}\alpha s}{(s+1)}\right]\equiv\left(\frac{1}{4\pi}\right)\left(\frac{2v}{3\Gamma_s}\right)\T{I}_{\delta\epsilon 1}.
\label{25}
\eea
From Eqs.~(\ref{gtnew})-(\ref{de}) the total energy and momentum deposition into the medium can be written as
\bea
\delta\epsilon(\mathbf{x},t)&=&\left(\frac{1}{4\pi}\right)\left(\frac{2v}{3\Gamma_s}\right)^2\left(\frac{9C_\kappa}{8}\right)\left[\kappa I_0^{\delta\epsilon}+\left(\frac{8}{9v}\right)\T{I}^{\delta\epsilon}\right]\nn
&\equiv&\left(\frac{1}{4\pi}\right)\left(\frac{2v}{3\Gamma_s}\right)^2\left(\frac{9}{8}\right)I^{\delta\epsilon}\left(\alpha,\beta,\kappa\right),
\label{finalenergy}
\eea
\bea
\n{g}(\mathbf{x},t)&=&\left(\frac{1}{4\pi}\right)\left(\frac{2v}{3\Gamma_s}\right)^2C_\kappa v\left[\kappa\n{I}_0^g+\T{\n{I}}^g\right]\nn
&\equiv&\left(\frac{1}{4\pi}\right)\left(\frac{2v}{3\Gamma_s}\right)^2v \n{I}^{g}\left(\alpha,\beta,\kappa\right).
\label{finalmomentum}
\eea

We now proceed to study how this momentum is distributed in transverse and longitudinal modes.

\section{Energy-momentum deposition}\label{III}

First, let us consider the case discussed in Ref.~\cite{ADY} where $\eta/s_0=1/4\pi$. When the range for the traveled path length $L$ is such that 7 fm $<L<$ 10 fm, then $75<\kappa<100$. Figure~\ref{superficies} shows the three dimensional plots for $I_{g_{Tz}}$ and $I_{g_{\delta\epsilon}}$ as functions of $\alpha$ and $\beta$ for $\kappa=75,\ 100$. The plots are shown in the range $0.5<\alpha <4$ and $-5<\beta <5$. The constant $C_\kappa$  in Eqs.~(\ref{finalenergy})-(\ref{finalmomentum}) is fixed by requiring that the total energy momentum deposited into the medium is the same as the case with a constant $dE/dx$, namely
\bea
\int d\alpha\ d\beta\left(\delta\epsilon^2+|\mathbf{g}|^2\right)=\int d\alpha\ d\beta\left(\delta\epsilon^2_0+|\mathbf{g}_0|^2\right).
\label{normalization}
\eea

Figure~\ref{I1} shows the integral defined in Eq.~(\ref{finalenergy}) and the different components of the integrals defined in Eq.~(\ref{finalmomentum}), integrated over the domain $\alpha_{min}=0.5 ,\ 1$, $\alpha_{max}=6$ and $-5 < \beta < 5$ for several values of $\kappa$. The figure also shows $I^{\delta\epsilon}_0$ and the components of $\n{I}^{g}_0$ which correspond to a constant energy-loss per unit length. Note that the hierarchy of momentum deposition is the same in both cases. This  means that the momentum is preferentially deposited also in the forward direction for this value of $\eta/s_0$.

The value  $\eta/s_0=1/4\pi$, corresponds to a universal lower bound for all relativistic quantum field theories in the strongly coupled limit~\cite{Neufeld79}. However, we can test the sensitivity to the momentum deposition when varying $\eta/s_0$~\cite{Neufeld79,etastong,arnoldeta, xueta}. Since $\Gamma_s$ is proportional to $\eta/s_0$, a traveled path length $L$ in the range 7 fm $<L<$ 10 fm, corresponds to different values of $\kappa$ than for the previously discussed case where we took $\eta /s=1/4\pi$. Note that since $\T{\delta\epsilon}$ is intrinsically negative, if we require that $\delta\epsilon = \kappa\delta\epsilon_0 - |\T{\delta\epsilon}|>0$ then not all values of $\kappa$ are allowed. To find a restriction involving $\kappa$ and $\eta/s_0$, note that
\bea
\kappa > \frac{|\T{\delta\epsilon}|}{\delta\epsilon_0}\approx 30
\label{restriction}
\eea
therefore
\bea
\frac{\eta}{s_0}<\frac{vt_{min}}{2}\frac{\delta\epsilon_0}{\T{\delta\epsilon}}T_0\approx 2.5\frac{1}{4\pi},
\label{restiction2}
\eea
where $t_{min}$ is the minimum time that we consider for the fast moving parton to have traveled in the medium. For definitiveness we take this parameter to be $t_{min}=7$ fm, given that the maximum time corresponds to twice the nuclear radius which for led nuclei is of order 10 fm.

\begin{figure*}[t]
{\centering
{\includegraphics[scale=0.45]{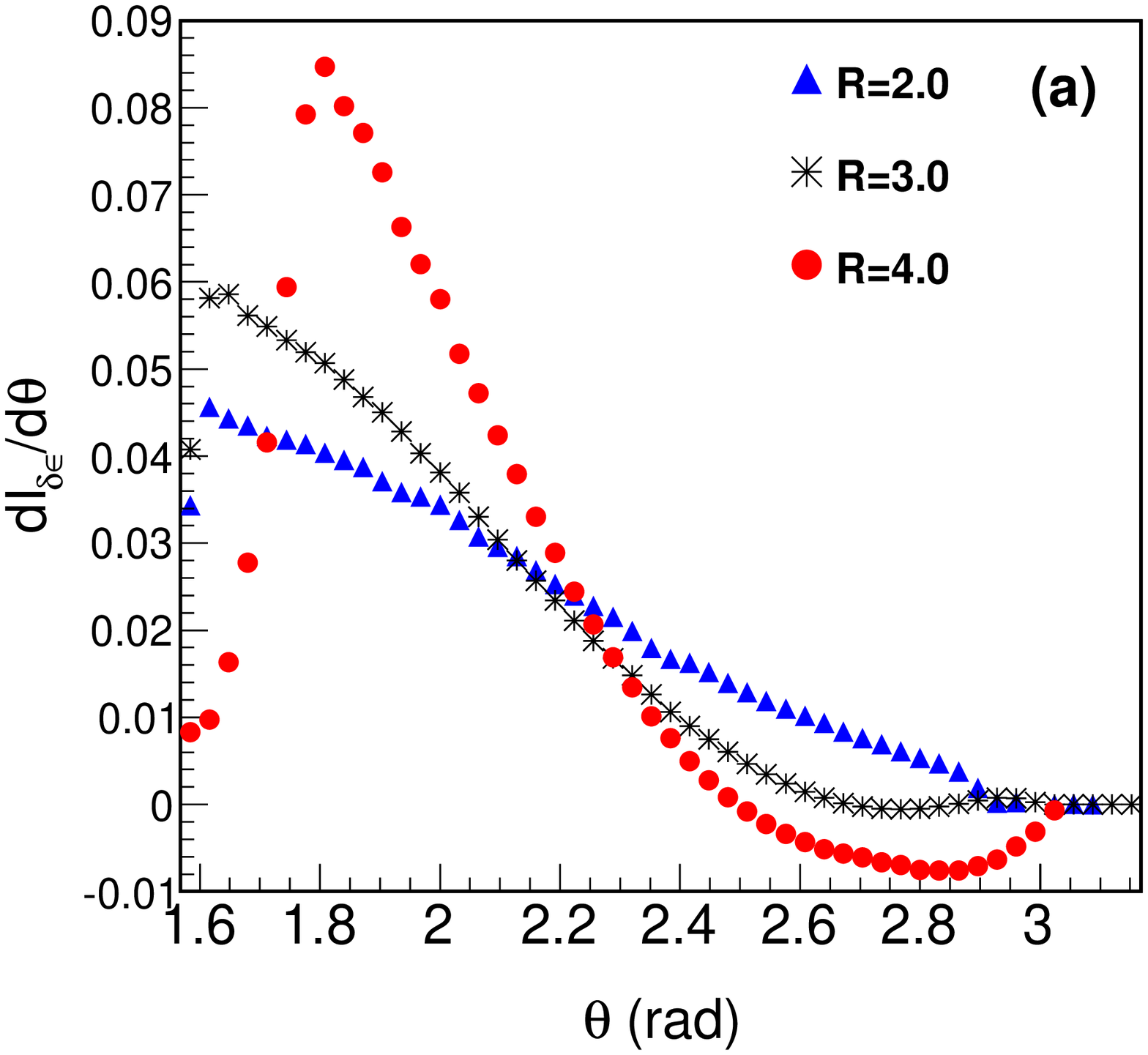}\includegraphics[scale=0.45]{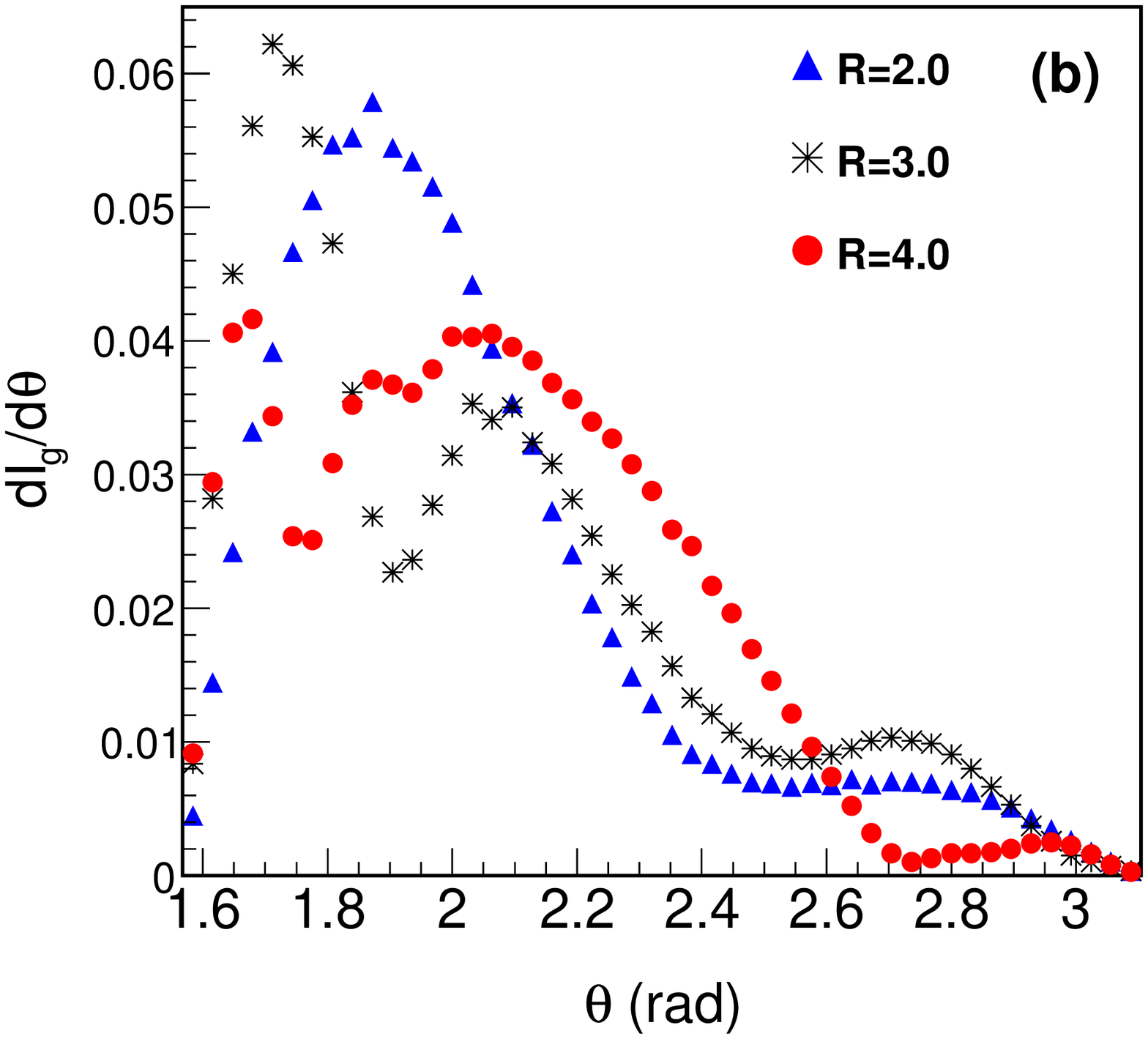}
}}
\caption{(Color online) Angular distribution of (a) energy density $dI_{\delta\epsilon}/d\theta$ and (b) momentum flux $d\n{I}_{g}/d\theta$ over an angular range $[\pi/2,\pi]$ at distances $R=2.0,\,3.0$ and $4.0$ in units of the sound attenuation length $\Gamma_s$ for $\eta/s_0=1/4\pi$ and $\kappa=75$.}
\label{flux}
\end{figure*}

Figure~\ref{superficies2} shows the integral defined in Eq.~(\ref{finalenergy}) and the component  $I_{g_{Tz}}$ of the integrals defined in Eq.~(\ref{finalmomentum}) as functions of $\alpha$ and $\beta$. The plots are shown in the range $0.5<\alpha <4$ and $-5<\beta <5$ for several values of $\kappa$. The normalization constant $C_\kappa$ is computed also from the requirement in Eq.~(\ref{normalization}). There is not much of a difference between the three dimensional surfaces in Fig.~\ref{superficies2} and those in Fig.~\ref{superficies}. This means that the spatial distribution of energy and momentum are very much alike for the cases with $\eta/s=1/4\pi$, $1.5/4\pi$. Figure~\ref{I2} shows the comparison between the integrals of Eqs.~(\ref{finalenergy}) and~(\ref{finalmomentum}), with respect to $\alpha$ and $\beta$, with the case corresponding to a constant energy-loss per unit length for $\eta/s=1.5/4\pi$. Note that the hierarchy of strengths for the momentum components for the case with $\eta/s=1.5/4\pi$ is maintained with respect to the case with $\eta/s=1/4\pi$. The only significant change comes from the energy deposition which is 30\% smaller in the latter case.

For completeness, we also study the case with $\eta/s=2/4\pi$. Figure~\ref{superficies3} shows the three dimensional plots corresponding to Eqs.~(\ref{finalenergy}) and~(\ref{finalmomentum}) for several values of $\kappa$. Figure~\ref{I3} shows the comparison between all components of these integrals and $I^{\delta\epsilon}_0$ and the components of $\n{I}^{g}_0$ which correspond to a constant energy-loss per unit length for the case $\eta/s=1/4\pi$. The normalization constant $C_\kappa$ is computed also with the requirement in Eq.~(\ref{normalization}). Note that the energy deposition decreases about 60\% with respect to the case with $\eta/s=1/4\pi$ but the hierarchy of strengths between the momentum modes remains the same as the case with $\eta/s=1/4\pi$.

In order to further study the energy-momentum deposition, we proceed as in Ref.~\cite{flux}, defining the energy density and momentum flux angular distributions as 
\bea
\frac{dI_{\delta\epsilon}}{d\theta}=2\pi R^2\sin\theta I_{\delta\epsilon},
\label{energyflux}
\eea
and
\bea
\frac{dI_{g}}{d\theta}&=&2\pi R^2\sin\theta\,\n{\hat{R}}\cdot\n{I}_g\nn
&=&2\pi R^2\sin\theta\left(|\n{g}_z|\cos\theta+|\n{g}_y|\sin\theta\right),
\label{momentumflux}
\eea
respectively, where $\n{R}$ is the distance vector from the source measured from the forward direction.

Figure~\ref{flux} shows the angular distribution for energy density (a) and momentum flux (b), for different values of distances to the source $R$ in units of sound attenuation length, for $\eta/s=1/4\pi$ and $\kappa =75$. Note that both angular distributions peak for angles close the source, which strengthens the conclusion that energy and momentum deposition is in the forward direction. The energy density increases and the momentum flux decreases with the distance to the source. This can be understood from the fact that the energy density contains an extra power of $R$ with respect to the momentum flux.

\section{Summary and conclusions}\label{IV}

In summary, we have studied the energy-momentum deposition produced by a fast moving parton traveling in a medium modeled by linear viscous hydrodynamics. The energy loss per unit length $dE/dx$ has been taken as proportional to the traveled length. We found that the transverse modes still dominate the momentum deposition and therefore this case is similar to the one where $dE/dx$ is taken as independent of the traveled path length. This situation is also maintained when the shear viscosity to entropy ratio $\eta/s_0$ is increased from its theoretical lower bound. The only significant change comes from the energy deposition which decreases as $\eta/s$ increases. Therefore, the momentum is forward peaked as in the case with constant energy loss per unit length as well as for the case of the lower value of $\eta/s$ previously studied~\cite{ADY}. We conclude that for the cases where $dE/dx$ is constant or proportional to the path length, as well as for larger than the lower theoretical bound values of $\eta/s$, the energy and momentum are preferentially deposited along the direction of motion of the traveling parton. Therefore, for the cases studied, the conical emission of particles is suppressed with respect to the forward emission, which means for instance that it is unlikely that the propagation of a single fast moving parton leads to the appearance of a double peak structure in azimuthal  angular correlations in heavy-ion collisions.

We also point out that it is easy to generalize the present studies to the case where $dE/dx\propto L^n$ with $n$ an integer larger than 1. In such a situation, the Fourier transform of $J^{\nu}(\mathbf{x},t)\propto \delta^{(n)}\left(\omega-\mathbf{k}\cdot\mathbf{v}\right)$, where $n$ is the nth-derivative of the delta function. Thus, the expressions for the energy and momentum deposition become polynomials of degree $n$ in $t$, where the coefficient of $t^n$ corresponds to the strength of the term $\mathbf{F}_0(\mathbf{x},t)$ which then becomes the dominant component for the range 7  fm $< L < 10$ fm. Therefore the hierarchy of the modes retain the general features already seen in the case of a constant energy loss per unit length profile.

\section*{Acknowledgments}

We acknowledge useful conversations with J. Jalilian-Marian during the initial stages of this work. Support for this work has been received in part from CONACyT-M\'exico under grant number 128534, from PAPIIT-UNAM under grant  number IN101515 and from {\it Programa de Intercambio UNAM-UNISON} and {\it Programa Anual de Cooperaci\'on Acad\'emica UAS-UNAM}. M. E. T.-Y. acknowledges support from the CONACyT-M\'exico sabbatical grant number 232946.

\end{document}